\documentclass[showpacs]{iopart}
\usepackage{amssymb,eucal,amsthm,epsf,epsfig,graphicx,float}
\usepackage{amsfonts}
\usepackage[dvips]{color}
\usepackage[dvips]{color}
\usepackage{setstack}
\usepackage[french,english]{babel}

\newcounter{parentequation}
\newenvironment{subeq}{\setcounter{parentequation}{\value{equation}}
\setcounter{equation}{0}
\addtocounter{parentequation}{1}
\renewcommand{\theequation}{\arabic{parentequation}\alph{equation}}}%
{\setcounter{equation}{\value{parentequation}}
\renewcommand{\theequation}{\arabic{equation}}}
\def\e{{\mathop{\,\rm e}}}
\def\ii{\dot{\textrm{\i}\hspace{-1.7pt}\textrm{\i}}}
\def\iii{\dot{\textrm{\i}\hspace{-1.5pt}\textrm{\i}}}
\def\R{\mathbb{R}}
\def\C{\mathbb{C}}

\def\N{\mathbb{N}}

\def\sinh{{\mathop{\rm sinh}}}

\newcommand{\largesim}[1]{\raisebox{-5pt}{\hbox{$
\widetilde{\scriptstyle#1}$}}}
\begin{document}
\title{The one-dimensional Coulomb Problem}
\author{G. Abramovici}
\ead{abramovici@lps.u-psud.fr}
\address{Laboratoire de Physique des Solides,
Univ. Paris Sud, CNRS, UMR 8502, F-91405 Orsay Cedex France}
\author{Y. Avishai}
\ead{yshai@bgu.ac.il}
\address{Department of Physics and Ilse Katz Center for Nanotechnology,\\
Ben-Gourion University, Beer-Shiva 84105, Israel \\
Hong Kong University of Science and Technology, Clear Water Bay, Kawloon, Hong Kong}
\begin{abstract}
One-dimensional scattering by a Coulomb potential $V(x)={\lambda\over|x|}$ is
studied for both repulsive ($c>0$) and attractive ($c<0$) cases. Two
methods of regularizing the singularity at $x=0$ are used, yielding the
same conclusion, namely, that the transmission vanishes. For an attractive
potential ($c<0$), two groups of bound states are found. The first one consists
of \emph{regular} (Rydberg) bound states, respecting standard orthogonality
relations. The second set consists of \emph{anomalous} bound states (in a sense
to be clarified), which always relax as coherent states.
\end{abstract}

%\maketitle
\section{Introduction}

One-dimensional quantum Hamiltonians are very useful for modeling simple
quantum systems. Beside their ubiquitous importance in the study of transmission
and tunneling experiments, numerous quantum systems in higher dimensions
can be reduced to one-dimensional ones, due to symmetry (for instance radial 
wave functions in a central potential) or specific physical properties
(Josephson junctions or edge states in the quantum Hall effect are just two
examples). 

The aim of the present work is to examine one-dimensional scattering 
by a three-dimensional coulomb potential
$V(x)={qq'\over4\pi\epsilon_{\rm o}|x|}$,  starting from the Schr\"odinger
equation with Hamiltonian $H={p^2\over2m}+V$, for an eigenstate $\psi(x)$, with
$x\in\R^\ast\equiv\R\setminus\{0\}$,
\begin{equation}
\label{SchrEqx}
-\frac{d^2\psi}{dx^2}(x)+\frac{\lambda}{|x|}\psi(x)=e \psi(x)\ ,
\end{equation} 
with $\lambda\!=\!{2mqq'\over4\pi\epsilon_{\rm o}\hbar^2}$ and
$e\!=\!{2mE\over\hbar^2}$ where $E$ is the energy. $\lambda>0$ corresponds to
the repulsive potential, $\lambda<0$ to the attractive one; The boundary
conditions will be specified later on. 
This is referred to as the \emph{one-dimensional Coulomb potential problem}.
Although it has recently been studied\cite{Mineev}, we find it useful to
analyze it using somewhat different approach. As it turns out, there are some
subtleties involved, which might affect some of the conclusions reached in
Ref.~\cite{Mineev}.

One of the main advantages encountered in the quantum Coulomb problem is
that the exact wave functions are computable. In three dimensions, it has been
shown eighty years ago\cite{Yost} that the asymptotic behavior of the wave
functions is somewhat distinct from that of plane waves. This property has been
shown to be valid also in one dimension\cite{Moshinsky}.

It proves useful to follow, first, the standard reduction of the Coulomb problem
in three dimensions into a radial one-dimensional equation, and to point out the
differences between this equation and Eq.~(\ref{SchrEqx}). Starting from the
three-dimensional Schr\"odinger  equation, carrying out partial wave expansion
$\Psi({\bf r})=\sum_{l=0}^{\infty} (2l+1)$\penalty-10000
$\psi_l(r) P_l(\cos \theta)$, and writing the radial wave function as
$\psi_{l}(r)=r^{-1}\phi_l(r)$, one obtains the radial Schr\"odinger equation for
$\phi_l(r)$, with $0<r<\infty$,
\begin{equation}
\label{SchrEqr}
\left[-\frac{d^2}{dr^2}+ \frac{l(l+1)}{r^2}+\frac{\lambda}{r} \right ]\phi_l(r)
=e \phi_l(r)\ .
\end{equation} 
For $l=0$ ($s$ wave scattering), Eq.~(\ref{SchrEqr}) has the same form as
Eq.~(\ref{SchrEqx}). The two basic solutions of Eq.~(\ref{SchrEqr}) are the 
regular one, satisfying $\phi_l(0)=0$, and the singular one, satisfying
$\phi_l(0)\ne0$. The singular solution should be discarded: if not, for $l>0$,
the probability of finding the particle in a sphere of radius $R$, 
$P_l(R)=\int_0^R\!\rho_l(r)2\pi r^2dr$ becomes infinite for
any $R$~; for $l=0$, the situation is more subtle, $P_0(R)$ remains finite,
but the radial current $J_0(R)=\int_0^R j_0(r)2\pi r^2dr>0$ becomes nonzero,
which is impossible for an $s$ state\cite{Basdevant,Shankar}. 

A couple of difficulties arise when Eq.~(\ref{SchrEqx}) is
considered as compared  with Eq.~(\ref{SchrEqr}):\begin{enumerate}
\item
The solutions of Eq.~(\ref{SchrEqx}) are required on $\R^\ast$, and not only on
its positive part $\R^\ast_+$. Note that $H$ is invariant under space inversion.
\item
The arguments used in the three-dimensional case to discard singular solutions
of Eq.~(\ref{SchrEqr})  are not valid\cite{Bohm} for the original problem
specified by Eq.~(\ref{SchrEqx}), and the imposition of scattering boundary
conditions requires their inclusion as well. The standard techniques used
for matching the wave function at $x=0$ require either the calculation of
$\psi'(\varepsilon)$ or of $\int_{-\varepsilon}^{\varepsilon}\! V(x) dx$ and
both quantities diverge logarithmically when $\varepsilon\to 0$. One must then
cope with ultraviolet divergences, which need to be regularized. 
\end{enumerate}
These difficulties lead us to the \emph{connection problem}, which can
be defined as follows:
Let us decompose Eq~.(\ref{SchrEqx}) into two equivalent coupled equations, one
defined on $\R^\ast_+$ with $\tilde V(x)=\frac{\lambda}{x}$, the general
solutions of which read 
\begin{subeq}
\begin{equation}
\psi_+(x)=A f(kx)+B g(kx)\ ,
\label{psip}
\end{equation}
and the second defined on $\R^\ast_-$  with $\tilde V(x)=-\frac{\lambda}{x}$,
the general solutions of which read 
\begin{equation}
\psi_-(x)=a \bar f(kx)+b \bar g(kx)\ .
\label{psim}
\end{equation}
\end{subeq}
Here, $f(x{>}0)$ and $\bar f(x{<}0)$ are regular solutions, while $g(x{>}0)$ and
$\bar g(x{<}0)$ are singular solutions, defined on the appropriate domains; the
relations between $f, g$ and ${\bar f}, {\bar g}$ will be clarified later on.
The connection problem consists in the calculation of the $2 \times 2$ matrix
expressing $(A,B)$ in terms of $(a,b)$. Since the derivative of the singular
solution diverges at $x=0$,  it is impossible to match  both $\psi$ and $\psi'$
at $x=0$. It is also not possible to use the method\cite{Tanaka,Lieb} employed
in a problem of scattering by a potential $V(x)=\lambda \delta(x)$ since the
latter potential is integrable at $x=0$,
$\int_{-\varepsilon}^{\varepsilon}\!V(x)dx=\lambda$, whereas the Coulomb
potential is not. Apparently, the connection problem cannot be solved in terms
of simple linear relations, and one needs to consider bilinear constraints (an
example of such a constraint is the current conservation $J(0^-)=J(O^+)$ around
$x=0$).

Our first task is to properly formulate and solve the scattering problem,
corresponding to $e>0$. To carry it out, we use two independent regularization
methods. One is based on bilinear constraints, which can be formulated in such
a way that ultraviolet divergences are canceled. The other method consists in
calculating the exact transmission for a truncated Coulomb potential
$V_\varepsilon$, with $V_\varepsilon(x)=0$ for $|x|<\varepsilon$,
$V_\varepsilon(x)=\lambda/|x|$ for $|x|>\varepsilon$ and letting
$\varepsilon\to 0$. With both methods, we arrive at the conclusion that the
transmission coefficient vanishes, $T=0$. The potential is perfectly reflective.
Moreover, this property of total reflection also holds for the attractive
potential ($\lambda<0$), whereas  classically the reflection vanishes; it is a
novel manifestation of perfect \emph{quantum reflection} from an attractive
potential. It is distinct from the standard example of quantum reflection from
an infinite attractive square well: in the latter case, the divergence of
$\int\! V(x) dx$ is faster than logarithmic, and the corresponding spectrum is
not bounded from below.

Our second goal is to calculate bound state energies and wave functions for 
an attractive potential ($\lambda<0$) (the one-dimensional ``hydrogen atom''
problem). The ensuing discrete part of the spectrum ($e<0$) appears to be rather
intriguing, as it is composed of two interlacing spectra. The first one
(reported also in Ref.~\cite{Mineev,Mineev2}) is the usual Rydberg spectrum,
with energies $E_n=-\frac{E_0}{n^2}$, with $n=1,2,\ldots$ The corresponding
wave functions are the \emph{regular} solutions of the differential
Eq.~(\ref{SchrEqx}). The energies of the second part of the spectrum
are shifted from the first ones through $n \to n+1/2$, that is, 
$\tilde E_n=-\frac{E_0}{(n+\frac{1}{2})^2}$, with $n=0,1,\ldots$ The
corresponding wave functions will be refereed to as \emph{anomalous} states, and
are constructed in terms of the singular solutions of Eq.~(\ref{SchrEqx}).
These solutions are square integrable but not orthogonal. A proper incorporation
of such states might require further insight into the basic principles of
quantum mechanics.

We organize the rest of the paper as follows: 
In section \ref{II}, we will first study the scattering problem, then explain,
in section~\ref{III}, the two regularization methods used to solve the
connection problem. The bound state problem will be analyzed in
section~\ref{IV}, where regular and anomalous states are introduced. Finally, a
short discussion of our results is carried in section~\ref{V}. 
Calculations requiring technical manipulations are collected in the appendices.
\section{The scattering problem }
\label{II}
\subsection{Scattering states}
\subsubsection{Basic solutions}
For the scattering problem, we have $e>0$ in Eq.~(\ref{SchrEqx}). It is
convenient to recast Eq.~(\ref{SchrEqx}) so that all quantities are
dimensionless. Let $k=\sqrt{e}$, $u=kx$,
$\eta=\lambda/(2k)={qq'\over4\pi\epsilon_{\rm o}\hbar}\sqrt{m\over2E}$ and
$\varphi(u)=\psi({u\over k})$. Then the equation for $\varphi$ is
\begin{equation}
\label{eq}
-{d^2\varphi\over du^2}(u)+2{\eta\over|u|}\varphi(u)=
\varphi(u)\ , \quad  u \in \R^\ast,
\end{equation}
with regular and singular solutions $f_\eta(u)$ and $g_\eta(u)$. Eq.~(\ref{eq})
is equivalent to the following couple of equations~: 
\begin{subeq}
\begin{eqnarray}
\label{eqpos}
-{d^2\varphi\over du^2}(u)+2{\eta\over u}\varphi(u)&=&
\varphi(u)\quad\hbox{for }u>0\ ;\\
-{d^2\varphi\over du^2}(u)-2{\eta\over u}\varphi(u)&=&
\varphi(u)\quad\hbox{for }u<0\ .
\label{eqneg}
\end{eqnarray}
\end{subeq}
The solutions of Eq.~(\ref{eqpos}) are known as Coulomb $s$ wave
functions~\cite{Yost,Abramowitz} with $L=0$. We will write $F_\eta(u)$ the
regular solution and $G_\eta(u)$ the singular (logarithmic) one:
\begin{subeq}
\begin{eqnarray}
\label{H1F1}
\!\!\!\!\!\!\!\!\!\!\!\!\!\!\!\!\!\!\!\!\!\!\!\!\!\!
F_\eta(u)&=&C_\eta u\e^{-\iii u}M(1-\ii \eta,2,2\ii u)\ ;
\\
\!\!\!\!\!\!\!\!\!\!\!\!\!\!\!\!\!\!\!\!\!\!\!\!\!\!
G_\eta(u)&=&\Re\left(2\eta{u\e^{-\iii u}\Gamma(-\ii\eta)\over C_\eta}
U(1-\ii \eta,2,2\ii u)\right)
\nonumber\\
\!\!\!\!\!\!\!\!\!\!\!\!\!\!\!\!\!\!\!\!\!\!\!\!\!\!
\!\!\!\!\!\!\!\!\!\!\!\!\!
&=&%\!\!\!\!\!\!\!\!
2\eta{u\e^{-\iii u}\Gamma(-\ii\eta)\over C_\eta}
U(1-\ii \eta,2,2\ii u)-\ii(-1+\pi\eta+2\iota_\eta)F_\eta(u)/C_\eta^2\ ,
\label{HU}
\end{eqnarray}
where
$$
C_\eta=\e^{-{\pi\eta\over2}}\sqrt{\pi\eta\over\sinh(\pi\eta)}\quad\hbox{and}
\quad\iota_\eta=\eta\Im(\Gamma(1-\ii\eta))\ .
 $$
In these equations, $M$ is the regular confluent hypergeometric function, also
written as $_1F_1$, and $U$ is the logarithmic (also called irregular) confluent
hypergeometric function\cite{attention}. Both $F_\eta$ and $G_\eta$ are
\emph{real}. Thus, the solutions of Eq.~(\ref{eq}) for $u>0$ are
$f_\eta(u)=F_\eta(u)$ and $g_\eta(u)=G_\eta(u)$, $\forall\eta$. 

Consider now the domain $u<0$. In principle, finding the solutions of
Eq.~(\ref{eqneg}) can be achieved by direct continuation of $F_\eta(u)$ and
$G_\eta(u)$. Practically, this requires some care, especially for $G_\eta$.
$F_\eta$ can be continued analytically since it is regular at $u=0$, while
for $G_\eta(u)$ one has to  avoid the divergence of $G'_\eta$ at $u=0$.
Since (\ref{eqpos}) is valid for any sign of $\eta$, we simply need to change
$\eta\to-\eta$ in the previous expressions, to get the solutions of
(\ref{eqneg}), thus we get $f_\eta(u)=F_{-\eta}(u)$ and
$g_\eta(u)=G_{-\eta}(u)$ $\forall u<0$ and $\forall\eta$. It should be pointed
out that, in the imaginary part of (\ref{HU}), the factor before $F_\eta$ does
not follow the $\eta\to-\eta$ transformation\cite{nonelucide}. The right
expression is (note that $C_{-\eta}=\e^{\pi\eta}C_\eta$), $\forall u<0$~:
\begin{equation}
\!\!\!\!\!\!\!\!\!\!\!\!\!\!\!\!\!\!\!\!\!\!\!\!\!\!
g_\eta(u)=-2\eta{u\e^{-\iii u}\Gamma(\ii\eta)\over C_{-\eta}}
U(1+\ii \eta,2,2\ii u)
-\ii(-1+\pi\eta+2\iota_\eta)F_{-\eta}(u)/C_{-\eta}^2\ .
\label{HUter}
\end{equation}
\end{subeq}
One should also note that
relations~(14.1.14) to (14.1.20) of \cite{Abramowitz} extend for $\rho<0$ as
soon as one replaces $\log(2\rho)$ by $\log(-2\rho)$ in (14.1.14).

\begin{figure}%[-h]
\centering
\includegraphics[width=6cm]{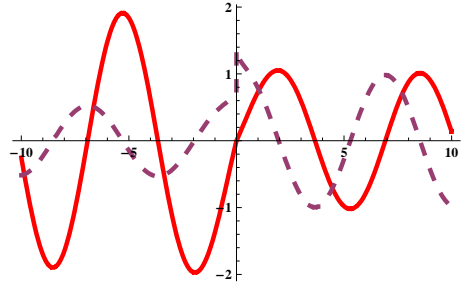}
\caption{$f_\eta$ (full line) and $g_\eta$ (dashed line) for
$\eta=1/5$.}
\label{courbep}
\end{figure}
Basic solutions $f_\eta(u)$ and $g_\eta(u)$ are defined on $\R^\ast$ and shown
on Fig.~\ref{courbep}. These solutions are constructed so that
Eqs.~(\ref{eqpos},\ref{eqneg}) are satisfied for both $u>0$ and $u<0$, yet 
the matching condition at $u=0$ is not addressed yet. This will be carried out
when we solve the connection problem.

\subsubsection{The general solution}
Having defined the \emph{basic solutions}, we can now form
the \emph{general solution} as  a linear combination of $f_\eta(u)$ and
$g_\eta(u)$, on each side of $u=0$. We use expressions (\ref{psip}) for
$u>0$ and (\ref{psim}) for $u<0$. Now, the relation between $f$ and $\bar f$ and
that between $g$ and $\bar g$ are well established, so that bar $\bar{ }\ $\ can
be omitted. With these notations, the general solution writes
\begin{equation}
\varphi(u,\eta)=
\cases{Af_\eta(u)+Bg_\eta(u)&for $u>0$ ;\cr
af_\eta(u)+bg_\eta(u)&for $u<0$ . \cr}
\end{equation}

The linearity of Schr\"odinger equation implies that the connection
problem eventually reduces in finding the $2\times2$ matrix $D$, which obeys
\begin{equation}
\left(\matrix{A\cr B\cr}\right)=
D\left(\matrix{a\cr b\cr}\right)
\quad~ \mbox{with} \quad~ \det(D)\ne0\ .
\label{connection}
\end{equation}

\subsubsection{Transfer matrix}

It should be stressed that $D$ is \emph{not} the transfer matrix ${\cal T}$
because ${\cal T}$ transforms incoming and outgoing (distorted) plane waves at
$u \to -\infty$ to those at $u \to \infty$. In order to identify these
asymptotic waves, we need first to examine the asymptotic behavior of the
function $\varphi(u,\eta)$ when $u \to \pm \infty$. 

The asymptotic behaviors of $F_\eta(u)$ and $G_\eta(u)$, for $u\to+\infty$,
have been established  a long time ago in Ref. \cite{Yost}:
\begin{subeq}
\begin{eqnarray}
\!\!\!\!\!\!\!\!\!\!\!\!\!\!\!\!\!\!
\!\!\!\!\!\!\!\!\!\!\!\!\!\!\!\!\!\!\!\!\!\!\!\!\!\!
\nonumber F_\eta(u)&=&\!\!\!\!\!
(1+{\eta\over2u}+{5\eta^2-\eta^4\over8u^2}{+}..)\sin(u-\Theta_\eta(u))
+\,({\eta^2\over2u}-{2\eta-4\eta^3\over8u^2}{+}..)\cos(u-\Theta_\eta(u))\\
\!\!\!\!\!\!\!\!\!\!\!\!\!\!\!\!\!\!
\!\!\!\!\!\!\!\!\!\!\!\!\!\!\!\!\!\!\!\!\!\!\!\!\!\!
&\largesim{u\to\infty}&\sin(u-\Theta_\eta(u))\ ;
\label{compFplus}\\
\!\!\!\!\!\!\!\!\!\!\!\!\!\!\!\!\!\!
\!\!\!\!\!\!\!\!\!\!\!\!\!\!\!\!\!\!\!\!\!\!\!\!\!\!
\nonumber G_\eta(u)&=&\!\!\!\!\!
(1+{\eta\over2u}+{5\eta^2-\eta^4\over8u^2}{+}..)\cos(u-\Theta_\eta(u))
-\,({\eta^2\over2u}-{2\eta-4\eta^3\over8u^2}{+}..)\sin(u-\Theta_\eta(u))\\
\!\!\!\!\!\!\!\!\!\!\!\!\!\!\!\!\!\!\!
\!\!\!\!\!\!\!\!\!\!\!\!\!\!\!\!\!\!\!\!\!\!\!\!\!\!
&\largesim{u\to\infty}&\cos(u-\Theta_\eta(u))\ ;
\label{compGplus}
\end{eqnarray}
\end{subeq}
with
\begin{equation} \label{thetaeta} 
\Theta_\eta(u)=\eta\log(2u)-\arg[\Gamma(1+\ii\eta)]\ .
\end{equation}

Derivation of the asymptotic behaviours of $F_\eta(u)$ and $G_\eta(u)$, for
$u\to -\infty$, is more subtle. Their determination (6c) and (6d) of
Ref.~\cite{Mineev} is to be reconsidered\cite{mistake}. In
Appendix~\ref{asymptotic}, we find
\begin{subeq}
\begin{eqnarray}
F_\eta(u)&=&\!\!\!\!\!\!\!
\e^{-\pi\eta}(1+{\eta\over2u}+{5\eta^2-\eta^4\over8u^2}{+}..)
\sin(u-\Theta_\eta(u))\nonumber\\
&&+\,\e^{-\pi\eta}({\eta^2\over2u}-{2\eta-4\eta^3\over8u^2}{+}..)
\cos(u-\Theta_\eta(u))\nonumber\\
&\largesim{u\to-\infty}&
\e^{-\pi\eta}\sin(u-\Theta_\eta(u))\ ;
\label{compFmoins}\\
G_\eta(u)&=&\!\!\!\!\!\!\!
\e^{\pi\eta}(1+{\eta\over2u}+{5\eta^2-\eta^4\over8u^2}{+}..)
\cos(u-\Theta_\eta(u))\nonumber\\
&&-\,\e^{\pi\eta}({\eta^2\over2u}-{2\eta-4\eta^3\over8u^2}{+}..)
\sin(u-\Theta_\eta(u))\nonumber\\
&\largesim{u\to-\infty}&
\e^{\pi\eta}\cos(u-\Theta_\eta(u))\ .
\label{compGmoins}
\end{eqnarray}
\end{subeq}
Thus, the asymptotic form of the solution $\varphi(u,\eta)$, is
\begin{subeq}
\begin{eqnarray}
\!\!\!\!\!\!\!\!\!\!\!\!\!\!\!\!\!\!\!\!\!\!\!\!\!\!\!\!\!\!\!\!\!\!
\varphi(u,\eta)&\largesim{u\to\infty}&
A\sin(u-\Theta_\eta(u))+B\cos(u-\Theta_\eta(u))
\nonumber\\
\!\!\!\!\!\!\!\!\!\!\!\!\!\!\!\!\!\!\!\!\!\!\!\!\!\!
\!\!\!\!\!\!\!\!\!\!\!\!\!\!\!\!\!\!\!\!\!\!\!\!\!\!
&&={B-\ii A\over2}\e^{\iii(u-\Theta_\eta(u))}
+{B+\ii A\over2}\e^{\iii(\Theta_\eta(u)-u)}\ ;
\label{pinfty}\\
\!\!\!\!\!\!\!\!\!\!\!\!\!\!\!\!\!\!\!\!\!\!\!\!\!\!\!\!\!\!\!\!\!\!
\nonumber\varphi(u,\eta)&\largesim{u\to-\infty}&
a\e^{\pi\eta}\sin(u+\Theta_\eta(u))+b\e^{-\pi\eta}\cos(u+\Theta_\eta(u))
\nonumber\\
\!\!\!\!\!\!\!\!\!\!\!\!\!\!\!\!\!\!\!\!\!\!\!\!\!\!
\!\!\!\!\!\!\!\!\!\!\!\!\!\!\!\!\!\!\!\!\!\!\!\!\!\!
&&={b\e^{-\pi\eta}-\ii a\e^{\pi\eta}\over2}\e^{\iii(u+\Theta_\eta(u))}
+{b\e^{-\pi\eta}+\ii a\e^{\pi\eta}\over2}\e^{-\iii(\Theta_\eta(u)+u)}\ .
\label{ninfty}
\end{eqnarray}
\end{subeq}
The transfer matrix ${\cal T}$ relates the coefficients of the distorted plane
waves at $u \to \infty$ with those at $u \to -\infty$:
\begin{equation} \label{Transfert}
\pmatrix{B-\ii A\cr B+\ii A \cr}={\cal T}
\pmatrix{ b\e^{-\pi\eta}-\ii a\e^{\pi\eta} \cr
b\e^{-\pi\eta}+\ii a\e^{\pi\eta} \cr}.
\end{equation}
Solution of the scattering problem is equivalent to elucidation of the transfer
matrix. 
\subsection{Scattering}
\subsubsection{Transmission and reflection amplitudes}

Alternatively, we define transmission $t$ and reflection $r$ amplitudes in
terms of a wave $\varphi_\alpha$ propagating from $-\infty$ ($\alpha=\rm L$), or
from $\infty$ ($\alpha=\rm R$). Explicitly,
$$
\varphi_{\rm L}(u,\eta)\raisebox{-2pt}{\Bigg\{}
\begin{tabular}{cl}
$\largesim{u\to-\infty}$&
$\e^{\iii(u+\Theta_\eta(u))}+r_{\rm L} \e^{-\iii(u+\Theta_\eta(u))}$ ;\\
\\
$\largesim{u\to\infty}$&$t_{\rm L} \e^{\iii(u-\Theta_\eta(u))}$ ;
\end{tabular}
 $$
and
$$
\varphi_{\rm R}(u,\eta)\raisebox{-2pt}{\Bigg\{}
\begin{tabular}{cl}
$\largesim{u\to\infty}$&
$\e^{-\iii(u-\Theta_\eta(u))}+r_{\rm R} \e^{\iii(u-\Theta_\eta(u))}$ ;\\
\\
$\largesim{u\to-\infty}$&$t_{\rm R} \e^{-\iii(u+\Theta_\eta(u))}$ .
\end{tabular}
 $$

Time reversal invariance implies $t_{\rm R}=t_{\rm L} \equiv t$ and reflection
symmetry $H(-x)=H(x)$ implies $r_{\rm R}=r_{\rm L} \equiv r$ (to demonstrate it
properly, one must note that, if $\varphi(u,\eta)$ is
a solution, $\varphi(-u,\eta)$ is another solution, \emph{a priori} independent
of the first one). Some useful relations expressing $A,B,a,b$ in terms of $t,r$
are given in Appendix~\ref{Texpression}. 

The corresponding transmission and reflection coefficients are
\begin{equation}
\label{defRT}
T=|t|^2 \ ,\quad R=|r|^2\ ,
\end{equation}
and fulfill  $R+T=1$ (see Eq. (\ref{RT1})). For $t\ne0$, it is instructive to
express the ratio of some coefficients $a,A$ in terms of $T$, once for
$\varphi_{\rm L}$, and once for $\varphi_{\rm R}$ (see
Appendix~\ref{Texpression}):
\begin{eqnarray*}
{a_{\rm L}\e^{\pi\eta}\over A_{\rm L}}=
\epsilon'-2\ii\epsilon\sqrt{{1\over T}-1}
&\Rightarrow&\left|{a_{\rm L}\e^{\pi\eta}\over A_{\rm L}}\right|
=\sqrt{{4\over T}-3}\ge1\ ;\\
{a_{\rm R}\e^{\pi\eta}\over A_{\rm R}}=
{1\over \epsilon'-2\ii\epsilon\sqrt{{1\over T}-1}}
&\Rightarrow&\left|{a_{\rm R}\e^{\pi\eta}\over A_{\rm R}}\right|=
{1\over\sqrt{{4\over T}-3}}\le1\ ;
\end{eqnarray*}
these inequalities become equalities only for $T=1$. This proves that the
symmetry between the regular and the singular part of a wave function
$\varphi$ which occurs at $x=\pm\infty$ is broken at $x=0$ and that connection
relations are not trivial (except for $T=1$ and also the special case $T=0$).

\subsubsection{The $S$ matrix}

The $S$ matrix is related\cite{Merzbacher,Imry} to $t$ and $r$ and writes
\begin{equation}
\label{Smat}
S=\left(\matrix{r&t\cr t&r\cr}\right)\ .
\end{equation}

Using the unitarity of the $S$ matrix, it is useful to parametrize its elements
in terms of the transmission coefficient $T$ and a couple of two independent
numbers $\epsilon, \epsilon'=\pm 1$. First, we get the parametrization
of all coefficients $A_{\rm L}$, ..., $b_{\rm R}$, which we give in
Appendix~\ref{Texpression}. Then, we can prove the representation
\begin{equation}
S=\left(\matrix{T-1+\ii\epsilon\epsilon'\sqrt{T-T^2}&
\epsilon'T+\ii\epsilon\sqrt{T-T^2}\cr\epsilon'T+\ii\epsilon\sqrt{T-T^2}&
T-1+\ii\epsilon\epsilon'\sqrt{T-T^2}\cr}\right)
\label{smat}
\end{equation}
which is unitary, as required. We stress that this representation is not
universal\cite{nonuniversalite}, namely, it is peculiar to the 
Coulomb scattering problem as discussed here. 

We are now in a position to examine the connection problem.

\section{The connection problem}
\label{III}

The connection problem  is to relate $A,B$ to $a,b$ either by finding matrix $D$
in Eq.~(\ref{connection}), or, equivalently, transfer matrix ${\cal T}$ in
Eq.~(\ref{Transfert}), or, equivalently, the $S$ matrix in Eq.~(\ref{Smat}).
Since $\partial\varphi\over\partial u$ diverges as $u\to0$, it is not
legitimate to use the continuity of $\varphi$ and
$\partial\varphi\over\partial u$ at $u=0$. Thus, the issue  of the connection
problem can not be handled in solving linear equations of the wave function, and
one must address bilinear relations, related either to conservation laws or
to certain constraints. In the following analysis,  the behaviors of
$f_\eta(u)$, $g_\eta(u)$ and of their derivatives, for $u\sim0$, are required:
they are studied in Appendix~\ref{mclaurin}.

\subsection{Conservation laws and other constraints}
\subsubsection{Continuity of $\rho_\eta$}

The simplest physical relation that provides a connection at $x=0$ is the
continuity of the density of probability, $\rho_\eta(u)=|\varphi(u,\eta)|^2$.
With relations (\ref{Continue},\ref{continue}), one gets
\begin{subeq}
\begin{equation}
|B|^2\e^{\pi\eta}=|b|^2\e^{-\pi\eta}\quad\iff\quad
\left|{B\over b}\right|=\e^{-\pi\eta}\ .
\label{continu}
\end{equation}

In Appendix~\ref{Texpression}, we show that this relation actually simplifies as
\begin{equation}
B=\epsilon'\e^{-\pi\eta}b\ .
\label{continu2}
\end{equation}
\end{subeq}
where $\epsilon'=\pm1$ (note that the case $\epsilon'=-1$ implies a violation of
the continuity of $\psi$).

\subsubsection{Current conservation}

The conservation of  current $j(x)=
-\Re\Big(\ii\overline{\psi(x)}{d\psi\over dx}(x)\Big)$ is equivalent to
the unitarity of the $S$ matrix which is already verified. Therefore, it does
not help for the resolution of the connection problem.

\subsubsection{Orthonormality of scattering states}

Since the complete set of scattering wave functions is known, it is in principle
possible to examine the consequence of generalized orthogonality  relations. Let
us write $\psi(x,E,\alpha)=\varphi_\alpha(kx,{\lambda\over2k})$ with
$\alpha=\rm R,L$ (wave functions coming from $+\infty$ or $-\infty$ have
degenerate energies), 
\begin{equation}
\label{ortho}
\int\!\!dx\;\overline{\psi(x,E_1,\alpha_1)}\psi(x,E_2,\alpha_2)=
\delta(k_1-k_2)P_{\alpha_1\alpha_2}
\end{equation}
where $P$ is an unitary $2\times2$ matrix in the (R,L) space.

In Appendix~\ref{orthonormalisation}, using relations (\ref{Al},\ref{Bl},%
\ref{al},\ref{bl},\ref{Ar},\ref{Br},\ref{ar},\ref{br}), (\ref{pinfty}) and
(\ref{ninfty}), we calculate\cite{implicite}
\begin{eqnarray} 
\!\!\!\!\!\!\!\!\!\!\!\!\!\!\!\!\!\!\!\!\!\!\!\!\!\!\!\!\!\!
\!\!\!\!\!\!\!\!\!\!\!
\lim_{L\to\infty}
\int_{-L}^L\overline{\psi(x,E_1,\alpha_1)}\psi(x,E_2,\alpha_2)dx&=&\!\!\!
\left[(1{+}{\sqrt{R(\eta_2)T(\eta_1)}{-}\sqrt{R(\eta_1)T(\eta_2)}\over2}{\cal Z}
\epsilon\epsilon'(1{+}\ii))\right.
\!\!\!\!\!\!\!\!\!\!\!\!\!\!\!\!\!\!\!\!\!\!\!\!
\nonumber\\
\!\!\!\!\!\!\!\!\!\!\!\!\!\!\!\!\!\!
\lefteqn{\times\delta(k_1-k_2)
+\left(-{R(\eta_1)+R(\eta_2)\over2}+
\epsilon\epsilon'{\sqrt{R(\eta_1)T(\eta_1)}+\sqrt{R(\eta_2)T(\eta_2)}\over2}
\right.}
\!\!\!\!\!\!\!\!\!\!\!\!\!\!\!\!\!\!\!\!\!\!\!\!
\nonumber\\
\!\!\!\!\!\!\!\!\!\!\!\!\!\!\!\!\!\!\!\!\!\!\!\!
\!\!\!\!\!\!\!\!\!\!\!\!\!\!\!\!\!\!\!\!
\lefteqn{+\left.\ii({T(\eta_1)-T(\eta_2)\over2}-\epsilon\epsilon'
{\sqrt{R(\eta_1)T(\eta_1)}-\sqrt{R(\eta_2)T(\eta_2)}\over2})\right)
\delta(k_1+k_2)+c\Big]\delta_{\alpha_1\alpha_2}\ ,}
\label{orthoL}
\end{eqnarray}
where $c$ is a constant and $\cal Z$ a complex number given by,
\begin{equation}
\!\!\!\!\!\!\!\!\!\!\!\!\!\!\!\!\!\!\!\!\!\!\!
\!\!\!\!\!\!\!\!\!\!\!
{\cal Z}=\sqrt{R(\eta_1)R(\eta_2)}+\sqrt{T(\eta_1)T(\eta_2)}
+\ii\epsilon\epsilon'(\sqrt{R(\eta_2)T(\eta_1)}-\sqrt{T(\eta_2)R(\eta_1)})\ .
\label{facteurz}
\end{equation}
Since $k_1,k_2>0$ here, we can drop $\delta(k_1+k_2)$ in Eq.~(\ref{orthoL}),
which is irrelevant\cite{precisiondelta}. The established result in
Eq.~(\ref{orthoL}) that $P=I_2$ reflects the orthogonality of left and right
moving states. Scattering states can be orthonormalized in the extended sense if
and only if $(\sqrt{R(\eta_2)T(\eta_1)}-\sqrt{R(\eta_1)T(\eta_2)}){\cal Z}=0$.
This yields $T(\eta_1)=T(\eta_2)$ or $T(\eta_i)\in\{0,1\}$. The second condition
is actually a particular case of the first one, since otherwise, one could find
some energy $E$ such that $T(\eta^+)=1-T(\eta^-)$, which induces a non physical
discontinuity; however, this argument will not be needed in the following.
Having $T$ independent of $E$ is already a very strong
result\cite{nonuniversalite}. Yet, in order to completely elucidate the
connection problem, we will now address another constraint.

\subsubsection{Hermiticity of the Hamiltonian}
A successful issue for the connection problem is given by analyzing the
hermiticity of Hamiltonian $H$. For $E_1\ne E_2$, we consider two wave functions
$\psi_1$~: $x\mapsto\psi(x,E_1)$ and $\psi_2$~: $x\mapsto\psi(x,E_2)$
(degeneracy is not relevant here, and $R,L$ indices can be omitted). Since $H$
is hermitian, the hermitian product of $|\psi_1\rangle$ with $H|\psi_2\rangle$
must be conjugate with the hermitian product of $|\psi_2\rangle$ with
$H|\psi_1\rangle$. Explicitly,
$$
\int\!\! dx\overline{\psi(x,E_1)}
\left[\!{-}{\partial^2\psi\over\partial x^2}(x,E_2){+}
{\lambda\over|x|}\psi(x,E_2)\!\right]\!\!
{=}\!\!\int\!\!
dx\!\!\left[\!{-}\overline{{\partial^2\psi\over\partial x^2}(x,E_1)}{+}
{\lambda\over|x|}\overline{\psi(x,E_1)}\right]\!\!\psi(x,E_2)
 $$
\vglue-10pt
$$
\iff
\int\!\!dx\;\overline{\psi(x,E_1)}{\partial^2\psi\over\partial x^2}(x,E_2)
-\overline{{\partial^2\psi\over\partial x^2}(x,E_1)}\psi(x,E_2)=0\ ,
 $$
so that
\begin{equation}
\left[-\overline{\psi(x,E_1)}{\partial\psi\over\partial x}(x,E_2)
+\overline{{\partial\psi\over\partial x}(x,E_1)}
\psi(x,E_2)\right]_{-\infty}^\infty=0\ .
\label{twocurrent}
\end{equation}
In Eq. (\ref{twocurrent}), we calculate the Cauchy principal value of the
left term, which writes, in terms of dimensionless variables and function
$\varphi$~:
\begin{subeq}
\begin{equation}
\lim_{L\to\infty}{\lambda\over2}
\left[-{\overline{\varphi(u,\eta_1)}\over\eta_2}
{\partial\varphi\over\partial u}(u,\eta_2)+
\overline{{\partial\varphi\over\partial u}(u,\eta_1)}
{\varphi(u,\eta_2)\over\eta_1}\right]_{-L}^L\!\!.
\label{j2inf}
\end{equation}
Since $-\overline{\varphi(u,\eta_1)}{\partial\varphi\over\partial u}(u,\eta_2)
+\overline{{\partial\varphi\over\partial u}(u,\eta_1)}\varphi(u,\eta_2)$ is
divergent at $u=0$, one must use regularized integral around zero. Hence
one should add the Cauchy principal value:
\begin{equation}
\lim_{\varepsilon\to 0^+}{\lambda\over2}
\left[{\overline{\varphi(u,\eta_1)}\over\eta_2}
{\partial\varphi\over\partial u}(u,\eta_2)
-\overline{{\partial\varphi\over\partial u}(u,\eta_1)}
{\varphi(u,\eta_2)\over\eta_1}\right]_{-\varepsilon}^\varepsilon
\label{j2u0}
\end{equation}
\end{subeq}
and Eq. (\ref{twocurrent}) writes (\ref{j2inf})+(\ref{j2u0})=0. The
contribution (\ref{j2inf}) is found to vanish when $L\to\infty$ (detailed
calculations, using relations~(\ref{pinfty}), (\ref{Ar}, \ref{Br}, \ref{ar},
\ref{br}), are given in Appendix~\ref{hermiticite}) so the net expression of
Eq.~(\ref{twocurrent}) is determined by (\ref{j2u0}) which yields
\begin{eqnarray*}
0&=&{\cal Z}\Big\{\epsilon\epsilon'
({C_{\eta_1}\over\eta_1 C_{\eta_2}}\sqrt{R(\eta_1)T(\eta_2)}-
{C_{\eta_2}\over\eta_2 C_{\eta_1}}\sqrt{T(\eta_1)R(\eta_2)})\\
&&+{2\over C_{\eta_1}C_{\eta_2}}\Re\Big(\Gamma(1{+}\ii\eta_2)-
\Gamma(1{+}\ii\eta_1)\Big)
\sqrt{T(\eta_1)T(\eta_2)}\Big\}\ .
\end{eqnarray*}
Employing relations (\ref{Al},\ref{Bl},\ref{al},\ref{bl}), we get the very same
equation. Note that $h_1(\eta_1,\eta_2)\equiv{C_{\eta_1}\over C_{\eta_2}}$,
$h_2(\eta_1,\eta_2)\equiv{C_{\eta_2}\over C_{\eta_1}}$
and $h_3(\eta_1,\eta_2)\equiv{1\over C_{\eta_1}C_{\eta_2}}$ are independent
two-variable functions. Indeed, let us assume a linear combination,
\begin{equation}
\gamma_1 h_1+\gamma_2 h_2+\gamma_3 h_3=0.
\label{relation}
\end{equation}
Since $\sqrt{x\over\sinh(x)}$ and $\sqrt{\sinh(x)\over x}$ are one-variable
independent functions, if one keeps $\eta_2$ constant and considers
Eq.~(\ref{relation}) as an equation of variable $\eta_1$, one gets $\gamma_1=0$;
if one keeps $\eta_1$ constant and considers Eq.~(\ref{relation}) as an equation
of variable $\eta_2$, one gets $\gamma_2=0$; thus, $\gamma_3=0$ and the 
independence of the three functions is proved.
Now $\cal Z$, defined in (\ref{facteurz}), can never vanish.  Hence one gets
$$
R(\eta_1)T(\eta_2)=0\ ;\ T(\eta_1)R(\eta_2)=0\ ;\ T(\eta_1)T(\eta_2)=0\ .
 $$
The first two equations imply $T=0,1$, and the last one simply implies $T=0$.
This eventually proves\cite{nonuniversalite} that, indeed, $T(\eta)=0$.

\subsection{Regularization by truncation of the potential}
Here we propose another approach, which gives the same result: the divergences
are regularized by a truncation of the potential.

\subsubsection{Truncated half-potential}
In order to avoid the use of Coulomb wave functions for negative argument 
we calculate transmission and reflection amplitudes for a right
half-barrier, defined for $x>0$, and then use reflection symmetry 
to calculate them for a mirror symmetric barrier, defined for $x<0$. 
Then left and right barriers are combined using a composition formula for the
$S$ matrix, as suggested for instance in Ref.~\cite{AL}.

\begin{figure}[!h]
\centering
\includegraphics[width=7cm]{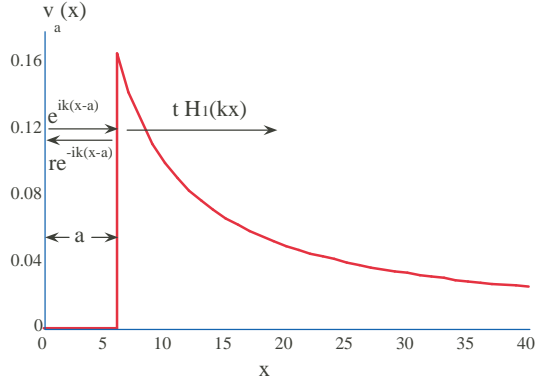}
\caption{Right-half truncated potential (\ref{potronq}) and wave function in the
two regions following Eq.~(\ref{psias}).}
\label{truncated_pot}
\end{figure}

The truncated right half-potential is, see Fig.~\ref{truncated_pot},
\begin{equation}
\label{potronq}
V_\varepsilon(x)=\cases{0&for $x\le\varepsilon$ ;\cr
\frac{\lambda}{x}&for $x>\varepsilon$ .\cr}
\end{equation}
and the Schr\"odinger equation with $V_\varepsilon(x)$ alone writes
$-\frac{d^{2}\psi(x)}{dx^{2}}+V_\varepsilon(x) \psi(x)=k^2 \psi(x)$~.

In order to avoid the $1/|x|$ singularity, the potential is assumed to be zero
for $0<x<\varepsilon$, but we have also performed our calculations with
$V_\varepsilon(x{<}\varepsilon)={\lambda\over\varepsilon}$, with no significant
changes. The cutoff parameter $\varepsilon>0$ is assumed small, and eventually
the limit $\varepsilon\to 0$ is taken on the sum of left and right barriers,
which corresponds to the complete Coulomb potential, since
\begin{equation}
{2m\over\hbar^2}V(x)=
\lim_{\varepsilon\to 0} V_\varepsilon(x)+V_\varepsilon(-x)\ . \label{vcx}
\end{equation} 

To calculate transmission and reflection amplitudes for the right barrier
consider a plane wave approaching the potential $V_\varepsilon$ from $-\infty$. It
is partially reflected by the barrier at $x=\varepsilon$, and the transmitted
wave is a Coulomb wave $tH\eta$, with $H_\eta(u)=F_\eta(u)+\ii G_\eta(u)$. Its
asymptotic behavior is
$$
H_\eta(u)\;\largesim{u\to \infty}\;\e^{\ii(u-\Theta_\eta(u))}\ .
 $$

The scattering boundary conditions for the wave function are, see
figure~\ref{truncated_pot},
\begin{equation}
\psi(x)=\cases{
e^{\ii k(x-\varepsilon)}+r e^{-\ii k(x-\varepsilon)}&for $x\le\varepsilon$\ ,\cr
t H_\eta(kx)&for $x>\varepsilon$\ .\cr}
\label{psias}
\end{equation}

We want to calculate reflection and transmission amplitudes $r$ and $t$ for this
right-half truncated Coulomb barrier $V_\varepsilon(x)$. Matching at
$x=\varepsilon$ yields
$$ 
1+r=t H_\eta(k\varepsilon)\ , \quad 1-r=-\ii t {\dot H}_\eta(k\varepsilon)\ ,
 $$
where $\dot H$ stands for $dH/du$, and thus
\begin{equation}
t=\frac{2}{H_\eta(k\varepsilon)-\ii  {\dot H}_\eta(k\varepsilon)}\ ;\quad 
r=\frac{H_\eta(k\varepsilon)+\ii  {\dot H}_\eta(k\varepsilon)}
{H_\eta(k\varepsilon)-\ii  {\dot H}_\eta(k\varepsilon)}\ .
\label{solhalf}
\end{equation}
In the limit $\varepsilon\to 0$, this implies
$$
t \to 0\ , \quad r \to -1\ . 
 $$
However, the limit $\varepsilon \to 0$ will not be taken here, but rather, at a later step. 

So far, we have considered transmission and reflection from the potential
$V_\varepsilon(x)$ where the incoming wave approaches the barrier from the left
region. If the wave would have come from the right, be partially transmitted to
the left and partially reflected back to the right, the transmission amplitude
would be the same, but the reflection would have a different phase. However,when
we combine the symmetric image of $V_\varepsilon(x)$ in order to account for the
Coulomb problem as asserted in Eq.~(\ref{vcx}), we employ the reflection
amplitude $r$, as a result of the analysis  developed in Ref.~\cite{AL}. This
procedure of combining the two barriers should be used \emph{before} the limit
$\varepsilon\to0$ is taken on Eqs.~(\ref{solhalf}). 
The transmission amplitude through the combined barrier 
$V_\varepsilon(x)+V_\varepsilon(-x)$ is
\begin{eqnarray}
\!\!\!\!\!\!\!\!\!\!\!\!\!\!\!\!\!\!\!\!\!\!\!\!
T_\varepsilon&=&\frac{t^2}{1-e^{2\ii k\varepsilon} r^2}\nonumber\\
\!\!\!\!\!\!\!\!\!\!\!\!\!\!\!\!\!\!\!\!\!\!\!\!
&=&
\label{Tfinal}
{4\over(1-e^{2\ii k\varepsilon})[H_\eta(k\varepsilon)^2-{\dot H}_\eta(k\varepsilon)^2)]
-2\ii(1+e^{2\ii k\varepsilon})H_\eta(k\varepsilon) {\dot H}_\eta(k\varepsilon)}
\ .
\end{eqnarray}
This formula is exact and expresses the transmission amplitude for a symmetric
combination of cutoff Coulomb barriers with a hole between $-\varepsilon$ and
$\varepsilon$. It uses Coulomb wave functions solely with positive argument.
Inspecting the two terms  of the denominator in Eq.~(\ref{Tfinal}), the first
term is found to vanish in the limit $\varepsilon\to0$, and hence:
$$
T_\varepsilon\approx \frac{\ii}{H_\eta(k\varepsilon){\dot H}_\eta(k\varepsilon)}
\mathop{\longrightarrow}\limits_{\varepsilon\to0} 0\ .
 $$
The upshot is that the transmission coefficient of combined left and right
barriers, which comprise Coulomb barrier as $\varepsilon\to0$, vanishes, that is
$T=\lim\limits_{\varepsilon\to0}T_\varepsilon=0$.

\subsubsection{A second form of truncated potential}

\begin{figure}%[-h]
\centering
\includegraphics[width=6cm]{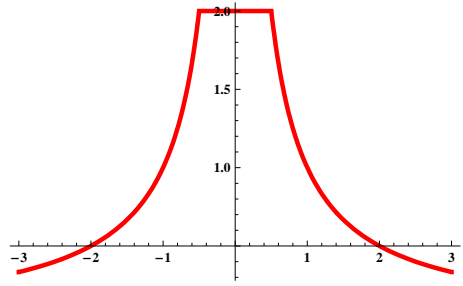}
\caption{Truncated potential for $\eta=1$, $\lambda=1$ and $\varepsilon=1$.}
\label{pottronq}
\end{figure}

We also considered a truncated potential $V_\varepsilon$, represented in
Fig.~\ref{pottronq} and defined as follow:
$\varepsilon>0$ and $\forall |x|\le\varepsilon$,
$V_\varepsilon(x)={\lambda\over\varepsilon}$, $\forall |x|>\varepsilon$,
$V_\varepsilon(x)={\lambda\over|x|}$. 

The transmission $T_\varepsilon$ can again be exactly calculated (the wave
function $\psi$ corresponding to given $(E,\varepsilon)$ and its derivative
$\psi'$ are continuous; we use first order Taylor expansion for the Coulomb wave
functions at connection points $x=\pm\varepsilon$). 

\begin{figure}%[H]
\centering
\includegraphics[width=7cm]{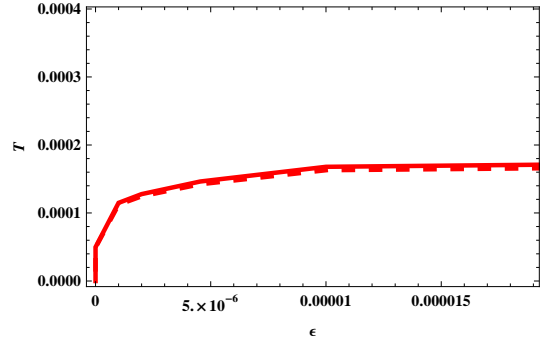}
\caption{Transmission $T_\varepsilon$ versus $\varepsilon$ in the repulsive
(plain line) or attractive (dashed line) case.}
\label{courb}
\end{figure}

One finds, in Fig.~\ref{courb} the curves of $T_\varepsilon$ versus
$\varepsilon$, for repulsive or attractive cases. We see that the transmission
$T_\varepsilon\to0$ as $\varepsilon\to0$. This confirms our analytical result.
We must precise that for some points of these figures, we used about $1000$
digit precision calculation, provided by a formal calculation with integers.

\section{Discrete spectrum : bound states}
\label{IV}

We come now to the case of an attractive potential, 
and look for bound states of negative energies. 
As is shown below, analytical expressions can be obtained for the
energies as well as for the wave functions\cite{Mineev3}.
\subsection{Analytical solutions}

For $e<0$, Eq. (\ref{eq}) is modified so that its right term writes
$-\varphi(u)$ instead of $\varphi(u)$. Note that $u=kx$ holds but now
$k=\sqrt{-e}$, since, for an  attractive potential, $\eta<0$. We will again
consider separately $u>0$ and $u<0$, and hence get the corresponding two
equations:
\begin{subeq}
\begin{eqnarray}
\label{lqpos}
-{d^2\varphi\over du^2}(u)+2{\eta\over u}\varphi(u)&=&
-\varphi(u)\quad\hbox{for }u>0\ ;\\
-{d^2\varphi\over du^2}(u)-2{\eta\over u}\varphi(u)&=&
-\varphi(u)\quad\hbox{for }u<0\ .
\label{lqneg}
\end{eqnarray}
\end{subeq}

In order to solve Eq. (\ref{lqpos}), we  need to generalize equations (14.1.6),
(14.1.14), (14.1.18), (14.1.19) and (14.1.20) of Ref.~\cite{Abramowitz} (for
$L=0$). This is carried out in Appendix~\ref{abramowitz}. Generalization of
(14.1.3) in Ref.~\cite{Abramowitz} is given below; relations (14.1.4),
(14.1.5), (14.1.15), (14.1.17) remain valid by construction. Incidentally, the
results of Appendix~\ref{abramowitz} can be regarded as a hyperbolic version of
the original relations in Ref.~\cite{Abramowitz}, since the solutions of
Eq.~(\ref{lqpos}) now read:
$$
J_\eta(u)\equiv u\e^{-u}M(1+\eta,2,2u)\ ,\ \qquad
K_\eta(u)\equiv 2u\e^{-u}U(1+\eta,2,2u)\ .
 $$
In analogy with the case of free states, the functions $J_{-\eta}$ and
$K_{-\eta}$ are solutions of (\ref{lqneg}) (the connection problem at $u=0$ will
be elucidated later on). A useful identity, which will be needed, is
\begin{equation}
\label{symJ}
J_{-\eta}(u)=-J_\eta(-u)\ .
\end{equation}
\subsection{Quantization}
For an arbitrary value of $\eta$, the solutions $J_\eta(u)$ and $K_\eta(u)$ of
Eq.~(\ref{lqpos}) diverge as $u \to \infty$ and the solutions $J_{-\eta}(u)$ and
$K_{-\eta}(u)$ of Eq.~(\ref{lqneg}) diverge as $u \to -\infty$. This is true for
almost all values of $\eta$, which therefore should be discarded as non
physical, except for a set of quantized values $\eta_n$ (equivalently
$e_n$ or $E_n$) such that $J_\eta(u>0)$ and $J_{-\eta}(u<0)$ are both square
integrable, and for another set of values $\tilde\eta_n$ (equivalently
$\tilde e_n$ or $\tilde E_n$) such that $K_\eta(u>0)$ and $K_{-\eta}(u<0)$ are
both square integrable. The complete spectrum, which is described below, is
composed of the union of set $\{E_n\}$, which is exactly Rydberg's spectrum, and
set $\{\tilde E_n\}$, the existence of which is indeed a surprise.

\subsubsection{The regular solutions}
Following the analysis of the hydrogen like atoms, it is verified that regular
solutions $J_\eta(u)$ and $J_{-\eta}(u)$ decay exponentially as $u\to\pm\infty$
only for a discrete set $\{\eta_n,\forall n\in\N^\star\}$ given by
\begin{equation}
\eta=\eta_n\equiv-n\ \iff\ 
E=E_n\equiv-{(qq')^2m\over2(4\pi\epsilon_{\rm o})^2\hbar^2n^2}\ .
\label{Eregul}
\end{equation}
The corresponding energies $E_n$ form the Rydberg spectrum of hydrogen like
atoms. In particular, the lowest energy is
$E_1=-{(qq')^2m\over2(4\pi\epsilon_{\rm o})^2\hbar^2}=-ZZ'E_I$, where $E_I$ is
the Rydberg energy. 

The question whether the set ${\eta_n}$ defined above can be used also for the
singular solutions is answered negatively, although the demonstration is not
immediate.  While $K_{-\eta_n}(u)$ diverges as  $u\to-\infty$, $K_{\eta_n}(u)$
\emph{does not diverge} as $u\to\infty$. Therefore, one may consider a mixed
solution $A J_{\eta_n}+B K_{\eta_n}$ for $u>0$ and $a J_{-\eta_n}$ for $u<0$.
However, as we shall see immediately below, $J_{\eta_n}(0)=J_{-\eta_n}(0)=0$,
while $K_{-n}(0^+)=1/C_{-\eta_n}$. Hence the continuity of the density 
$\rho$ at $x=0$ implies here $|B|=0$, which proves that a combination of
regular and singular solutions is not an eigenstate.

So far we have asserted the exponential decay of $J_{\pm\eta_n}$ as
$u\to\pm\infty$. The complete regular solutions $\forall n\in\N^\ast$ can be
constructed as $\zeta_n(u)=J_{\eta_n}(u)$ $\forall u>0$ and
$\zeta_n(u)=-\mu J_{\eta_n}(-u)$ $\forall u<0$, with $\mu\in\C$, (due 
to Eq.~(\ref{symJ}) and the reflection symmetry between Eqs.~(\ref{lqpos})
and (\ref{lqneg})). Explicitly (cf. Eq. (13.6.9) of Ref.~\cite{Abramowitz}),
\begin{equation}
\zeta_n(u)=-{u\over n}\e^{-|u|} L'_n(2|u|)
\cases{1&for $u>0$ ,\cr\mu&for $u<0$ ,\cr}
\end{equation}
where $L_n(z)$ is the Laguerre polynomial of order $n$, and $L_n'(z)=\frac{d L_n(z)}{d z}$. It will be shown below that $\mu=\pm1$.

The orthogonality and normalization of the corresponding wave functions
$\psi(x,E_n)=\zeta_n({\lambda x\over2\eta_n})=\zeta_n({|\lambda|x\over2n})$ can
be inspected by carrying out integration on the positive semi axis $\R_+$. Thus,
for the normalization we have,
$$
\int_0^\infty\!\!\!\!\!dx\;|\psi(x,E_n)|^2
=\int_0^\infty\!\!\!\!\!dx\;|\zeta_n(kx)|^2
={1\over k}\int_0^\infty\!\!\!\!\!\!du\;|\zeta_n(u)|^2
={2n\over|\lambda|}{n\over4}={n^2\over2|\lambda|}\ ;
 $$
which, with $|\mu|=1$, requires a  normalization factor equal to
$\sqrt{|\lambda|}\over n$; while for the orthogonality we find, 
$$
\int_0^\infty\!\!\!\!\!dx\;\overline{\psi(x,E_n)}\psi(x,E_{n'})
=\int_0^\infty\!\!\!\!\!
dx\;\overline{\zeta_n({|\lambda|x\over2n})}\zeta_{n'}({|\lambda|x\over2n'})
=0\quad \forall n\ne n'\ ,
 $$
due to orthogonality relations between Laguerre polynomials. 
 
\subsubsection{Anomalous solutions}
Quite remarkably, the anomalous solutions $K_\eta(u)$ and $K_{-\eta}(u)$ both
decay exponentially as $u\to\pm\infty$ only for a discrete set
$\{\tilde\eta_n,\forall n\in\N\}$ given by
\begin{equation}
\!\!\!\!\!\!\!\!\!\!\!\!\!\!\!\!\!\!\!\!\!\!\!\!\!\!\!\!
\eta=\tilde\eta_n=-n-{1\over2}\ \iff\ 
E=\tilde E_n
\equiv E_{n+{1\over2}}
=-{(qq')^2m\over2(4\pi\epsilon_{\rm o})^2\hbar^2(n+{1\over2})^2}\ ,
\label{Eanom}
\end{equation}
where $E_n$ is that of Eq. (\ref{Eregul}). The corresponding energies
$\tilde E_n$ form a separate spectrum interlacing the Rydberg one. From
Eq.~(\ref{Eanom}), one notes that $\tilde E_n={p^2\over(n+{1\over2})^2}E_p$,
$\forall p\in\N^\ast$, so that the minimum $\tilde E_0$ is lower than $E_1$ by a
factor of 4.

Note that, for $\eta\ne\tilde\eta_n$, $K_{-\eta}(u)$ is diverging
exponentially for $u\to-\infty$, while $K_\eta(u)$ does not diverge
for $u\to\infty$. Therefore, one should examine the possibility of a continuous
spectrum, by constructing a solution $A K_\eta(u)$ for $u>0$ and zero for $u<0$
for any such $\eta\ne\tilde\eta_n$; however, one can calculate
$K_\eta(0^+)=1/C_\eta\ne0$ for all $\eta<0$, so the continuity of the density
$\rho$ at $x=0$ implies $A=0$. This possibility is eventually discarded.
 
So far we have asserted the exponential decay of $K_{\pm\tilde\eta_n}$ as
$u\to\pm\infty$. In order to construct the complete anomalous solutions, one
needs to examine first the properties of $K_{-{\tilde \eta}_n}(u)$ for $u<0$ and
$n\in\N$. The imaginary part writes
$$
\Im(K_{-\tilde\eta_n}(u))=
{\sqrt{\pi}\over\gamma_n}J_{\tilde\eta_n}(u)\qquad\hbox{with }\ 
\gamma_n=(2n-1)!!/2^{n+1}\ ;
 $$
while, for the real part, there is a relation analogous to (\ref{symJ})~:
\begin{equation}
\label{symK}
K_{\tilde\eta_n}(-u)-
\ii{\sqrt{\pi}\over\gamma_n}J_{\tilde\eta_n}(-u)
=\nu_n K_{-\tilde\eta_n}(u)\qquad\forall u>0
\end{equation}
where $\nu_n=2^{2n+1}/((2n+1)(2n-1)!!)^2)$~: $K_{\tilde\eta_n}$ has even parity
(whereas $J_{\eta_n}$ has odd parity) if one omits rescaling factor $\nu_n$.

The complete anomalous solutions $\forall n\in\N$ can then be defined as
$\xi_n(u)=K_{\tilde\eta_n}(u)$ for $u>0$ and $\xi_n(u)=\nu K_{\tilde\eta_n}(-u)$
for $u<0$, due to Eq.~(\ref{symK}) and the reflection symmetry between
Eqs.~(\ref{lqpos}) and (\ref{lqneg}). It is not necessary to include the
factor $\nu_n$ here, since it is accounted for by the coefficient $\nu$.
The latter will be shown below to be $\nu=\pm 1$. In
Appendix~\ref{demonstration}, we prove that the anomalous solutions are
explicitly given by
\begin{equation}
\!\!\!\!\!\!\!\!\!\!\!\!\!\!\!\!\!\!\!\!\!\!\!\!\!\!\!
\xi_n(u)=
(p_n(|u|){\bf K}_0(|u|)+q_n(|u|){\bf K}_1(|u|))
{|u|\over(-2)^n\sqrt{\pi}}
\times\cases{1&for $u>0$ ,\cr \nu&for $u<0$ ,\cr}
\end{equation}
where polynomials $p_n(x)$ and $q_n(x)$ follow recurrence Eqs.~(\ref{recp}) and
(\ref{recq}), and $\bf K_n$ are the Bessel functions of the second kind. For
instance, $p_0=q_0=1$, $p_1(x)={3-4x}$, $q_1(x)={1-4x}$, $p_2(x)=4x(4x-9)+15$
and $p_2(x)=4x(4x-7)+3$ (more generally, these polynomials are proved to be real
with integer coefficients in Appendix~\ref{demonstration}). We are unaware of
any occurrence of this family of polynomials, which are worth being studied
further.

As for determining the constant $\nu$, contrary to the regular case,
$\xi_n(0)\ne0$. Hence, from the continuity of the density $\rho$, we deduce that
$$
\left|\xi_n(0^-)\right|=\left|\xi_n(0^+)\right|
 $$
in  analogy with Eq. (\ref{continu}). This implies $\nu=\pm1$ (we are studying
real solutions). Thus,  the anomalous solution $\xi_n$ is even for $\nu=1$ and
odd for $\nu=-1$.

Similarly to the case of regular solutions, the orthogonality and normalization
of the corresponding wave functions $\psi(x,\tilde E_n)=
\xi_n({\lambda x\over2\tilde\eta_n})=\xi_n({|\lambda|x\over2n+1})$ can be
inspected by carrying out integration on the positive semi axis $\R_+$. Thus,
for the normalization we have,
$$
\int_0^\infty\!\!\!\!\! dx|\psi(x,\tilde E_n)|^2=
\int_0^\infty\!\!\!\!\! dx|\xi_n(kx)|^2
={1\over k}\int_0^\infty\!\!\!\!\! du|\xi_n(u)|^2
={1\over|\lambda|}
\left({(2n+1)\beta_n\over2^{2n+2}\pi}+{\nu_n\pi\over2^{n+3}}\right)
 $$
The first coefficients $\beta_n$ can be easily computed, $\beta_0=3$,
$\beta_1=41$, $\beta_2=1063$. For large $n$, $\beta_n\sim5(2n+1)!!$.
Since we proved $\nu=\pm1$, one can deduce the exact normalization factor.

Strikingly, the anomalous solutions are not orthogonal to each other. As a
counter example, consider three hermitian products between anomalous states
$\xi_n$ and $\xi_p$ with $(n,p)=(0,1)$, $(0,2)$ and $(1,2)$, on the semi-axis
$\R_+$~:
\begin{eqnarray*}
\!\!\!\!\!\!\!\!\!\!\!\!\!\!\!\!\!\!\!\!\!\!\!\!\!\!\!
\!\!\!\!\!\!\!\!\!\!\!\!\!\!
\int_0^\infty\!\!\!\!\! dx\overline{\psi(x,\tilde E_0)}\psi(x,\tilde E_1)&=&
\int_0^\infty\!\!\!\!\! dx\overline{\xi_0({|\lambda|x})}
\xi_1({|\lambda|x\over3})\\
\!\!\!\!\!\!\!\!\!\!\!\!\!\!\!\!\!\!\!\!\!\!\!\!\!\!\!
\!\!\!\!\!\!\!\!\!\!\!\!\!\!\!\!\!\!\!\!\!\!\!\!\!\!\!
&=&{2\over|\lambda|}(
{3\over8\pi}-{9({\sf E}(-8)-3{\sf E}({8\over9})-3{\sf K}(-8)+
{\sf K}({8\over9}))+3\ln(729)\over64})\\
\!\!\!\!\!\!\!\!\!\!\!\!\!\!\!\!\!\!\!\!\!\!\!\!\!\!\!
\!\!\!\!\!\!\!\!\!\!\!\!\!\!\!\!\!\!\!\!\!\!\!\!\!\!\!
&\simeq&{2\over|\lambda|}0.0210133
\end{eqnarray*}
\begin{eqnarray*}
\!\!\!\!\!\!\!\!\!\!\!\!\!\!\!\!\!\!\!\!\!\!\!
\!\!\!\!\!\!\!\!\!\!\!\!\!\!\!\!\!
\int_0^\infty\!\!\!\!\! dx\overline{\psi(x,\tilde E_0)}\psi(x,\tilde E_2)&=&
\int_0^\infty\!\!\!\!\! dx\overline{\xi_0({|\lambda|x})}
\xi_2({|\lambda|x\over5})\\
\!\!\!\!\!\!\!\!\!\!\!\!\!\!\!\!\!\!
\lefteqn{={2\over|\lambda|}(
-{35\over48\pi}+{175({\sf E}(-24)-5{\sf E}({24\over25})-4{\sf K}(-24)
+{4\over5}{\sf K}({24\over25}))+27\ln(5)\over1728})}\\
\!\!\!\!\!\!\!\!\!\!\!\!\!\!\!\!\!\!\!\!\!\!\!
\!\!\!\!\!\!\!\!\!\!\!\!\!\!\!\!\!
\!\!\!\!\!\!\!\!\!\!\!\!\!\!\!\!\!\!\!\!\!\!\!\!\!\!\!
&\simeq&-{2\over|\lambda|}0.0319898
\end{eqnarray*}
\begin{eqnarray*}
\!\!\!\!\!\!\!\!\!\!\!\!\!
\!\!\!\!\!\!\!\!\!\!\!\!\!\!\!\!\!\!\!\!\!\!\!\!\!\!\!
\int_0^\infty\!\!\!\!\! dx\overline{\psi(x,\tilde E_1)}\psi(x,\tilde E_2)&=&
\int_0^\infty\!\!\!\!\! dx\overline{\xi_1({|\lambda|x\over3})}
\xi_2({|\lambda|x\over5})\\
\!\!\!\!\!\!\!\!\!\!\!\!\!\!\!\!\!\!\!\!\!\!\!\!\!\!\!
\!\!\!\!\!\!\!\!\!\!\!\!
\lefteqn{={2\over|\lambda|}(
{45\over32\pi}-{45(2705(3{\sf E}(-{16\over9})-5{\sf E}({16\over25}))
-2877(5{\sf K}(-{16\over9})-3{\sf K}({16\over25}))+15\ln(729))\over256})}\\
\!\!\!\!\!\!\!\!\!\!\!\!\!
\!\!\!\!\!\!\!\!\!\!\!\!\!\!\!\!\!\!\!\!\!\!\!\!\!\!\!
\!\!\!\!\!\!\!\!\!\!\!\!\!\!\!\!\!\!\!\!\!\!\!\!\!\!\!
&\simeq&{2\over|\lambda|}0.0188906
\end{eqnarray*}
where \textsf{K} is the complete elliptic integral of the first kind and
\textsf{E} is the complete elliptic integral of the second kind. It might be
argued that these integrals were calculated on the semi-axis $\R_+$, while the
hermitian product should be calculated on $\R$ and might vanish by symmetry
cancellation (in case of odd parity, integrals on $\R_+$ and on $\R_-$ have
opposite sign). However, since we have already proved that all anomalous wave
functions are either even or odd, then out of the three states
($\xi_0$, $\xi_1$, $\xi_2$), two have necessarily the same parity; thus,
the corresponding scalar product is non zero, and these solutions are not
orthogonal to each other.

This is a surprising result which requires more insight into the properties of
wave functions in quantum mechanics, which we will discuss briefly afterward.

\subsubsection{Orthogonality between regular and anomalous solutions}
Regular and anomalous solutions have different energies so they are expected to
be mutually orthogonal as well (see also the discussion afterward).

Performing the hermitian product on the semi-axis $\R_+$ of
$\psi(x,E_n)=\zeta_n({|\lambda|x\over2n})$ with
$\psi(x,\tilde E_p)=\xi_p({|\lambda|x\over2p+1})$ yields a non zero result. For
instance,
$$
\int_0^\infty\!\!\!\!\! dx\overline{\psi(x,\tilde E_0)}\psi(x,E_1)=
\int_0^\infty\!\!\!\!\! dx\overline{\xi_0({|\lambda|x})}
\zeta_1({|\lambda|x\over2})={2\over3\sqrt{\pi}|\lambda|}\ ;
 $$
similar expressions can be obtained for all $n\in\N^\ast$ and $p\in\N$, they can
all be written as  $r/(q\sqrt{\pi}|\lambda|)$, with integers $r$ and $q$
depending on $p$ and $n$. Thus, orthogonality between regular and anomalous wave
functions can be assured only by \emph{symmetry cancellation} of the right part
of the hermitian product (on $\R_+$) with its left part (on $\R_-$).
 
This leads to the following constraints: first, like the anomalous solutions, 
all regular solutions must have a definite parity. This is satisfied for
$\mu=\pm1$. Second, all regular solutions must have the same parity, and all
anomalous solutions must have the other parity. This means $\mu=\nu$ is fixed.
There remains a global choice of sign; either one chooses all regular solutions
to be odd and all anomalous solutions to be even or {\it vice versa}.

\begin{figure}%[-h]
\centering
\includegraphics[width=8cm]{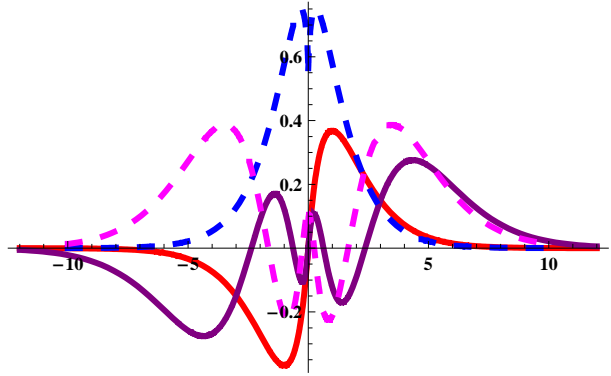}
\caption{$\zeta_1$ (red, full line), $\zeta_3$ (purple, full line),
$\xi_0$ (blue, dashed line) and $\xi_2$ (magenta, dashed line).}
\label{reprliees}
\end{figure}

While we have no rigorous argument for either case, one notes that the choice
$\mu=\nu=1$ implies that $\zeta_n$, $\zeta_n'$  and $\xi_n$ are continuous. This
seems to us the natural choice. Consequently,  regular solutions $\zeta_n$ are
odd and anomalous solutions $\xi_n$ are even. The first few solutions are shown
in Fig.~\ref{reprliees}.
With this choice, all solutions are continuous at $u=0$, whereas the first and
second derivative of $\xi_n$ are infinite at $u=0$ (this point is
actually a ramification point). 

\section{Discussion}
\label{V}

Despite its apparent simplicity, this one-dimensional problem leads to many
interesting results, some of them unexpected. In the following, we will list our
main results and discuss some of them.

\subsection{Zero transmission through the barrier}

The fact that $T=0$ for a repulsive infinite potential is in agreement with
classical mechanics. On  the contrary, for an attractive potential, it
contradicts classical mechanics. An example of perfect reflection from an
attractive potential, called quantum reflection, is provided by the infinite
square well potential:
$$
V(x)=V_{\rm o}\times
\cases{1&for $|x|\le a$ ,\cr0&for $|x|>a$ ,\cr}
\qquad V_{\rm o}\to-\infty\ ,
 $$
where $2a$ is the width of the well. The Coulomb potential provides us with a
new example of pure reflection. It differs from the infinite square well case by
the width, which becomes narrower as one goes down in energy, and by the
divergence of $\int V(x)dx$, which is logarithmic, while it is faster for the
square well potential. Note that both the Coulomb potential and the infinite
square well have an infinite number of bound states at negative energy. However,
while the spectrum of the former is bounded from below, the spectrum of the
latter is not. This is the only example of zero transmission and bounded
spectrum that we know of.

As a consequence of $T=0$, singular unbound wave functions are eventually
discarded, but the demonstration is much more involved than in the
three-dimensional case of Eq.~(\ref{SchrEqr}). If one looks back at relations
(\ref{Al},\ref{Bl},\ref{al},\ref{bl},\ref{Ar},\ref{Br},\ref{ar},\ref{br}), one
finds that all $B$ and $b$ coefficients cancel: the logarithmic solution is
completely suppressed, and therefore, the probability density is strictly zero
at $x=0$. In the case of $\psi_{\rm L}$, it is zero for $x\ge0$; in the case of
$\psi_{\rm R}$, it is zero for $x\le0$; the reflection process takes place
entirely on the half-line. This suppression at $x=0$ can be physically
interpreted as a hard wall repulsion. It is also true for regular bound states,
which probability density cancels at $x=0$. But it is not the case for anomalous
bound states, which show, here again, a special behavior.

\subsection{Representation of the $S$ matrix}

In one-dimensional scattering problem with symmetric potential $V(x)=V(-x)$, the
$S$ matrix is given by Eq.~(\ref{Smat}). Writing $t=\sqrt{T}\e^{\iii \theta_t}$
and $r=\sqrt{1-T}\e^{\iii \theta_r}$, the unitarity of $S$ implies
$\cos(\theta_t-\theta_r)=0$. Therefore, the most general expression of the $S$
matrix can reduce to
\begin{equation}
S=\e^{\iii \theta_t}\left(\matrix{\epsilon''\sqrt{1-T}&\sqrt{T}\cr
\sqrt{T}&\epsilon''\sqrt{1-T}\cr}\right)\ ,
\end{equation}
where $0\le\theta_t<2\pi$ and $\epsilon''=\pm1$ is an arbitrary sign. The form
of Eq.~(\ref{smat}) is unique to the Coulomb problem and reduces the number of
free parameters, since there is only one continuous parameter $T$ and two
arbitrary signs $\epsilon$ and $\epsilon'$. In particular, the phase $\theta_t$
is now given by
$$
\tan(\theta_t)={\epsilon\over\epsilon'\sqrt{T(1-T}}\ .
 $$
More precisely, (\ref{smat}) relies on relations~(\ref{pinfty}), (\ref{ninfty})
and on the reflection symmetry of the potential. For any symmetrical potential,
one can choose a basis of solutions $(f,g)$ such that (\ref{pinfty}) holds;
however, any generalization of relation (\ref{continu}) will fix the ratio $A/a$
or $B/b$ so that (\ref{ninfty}) will be changed.

Note that, with $T=0$, one simply gets $S=-I_2$.

\subsection{Non hermiticity of $H$}

The non orthogonality between anomalous bound states implies that $H$ is not
perfectly hermitian, because it is well established that the eigenstates of an
hermitian operator \emph{are} orthogonal. This problem is raised by the same
singularity than that, which is calculated in (\ref{j2u0}). Indeed, the quantity
$\Delta_{np}$ defined by
\begin{eqnarray*}
\!\!\!\!\!\!\!\!\!\!\!\!\!\!\!\!\!\!\!\!\!\!\!\!\!\!\!
\!\!\!\!\!\!\!\!\!\!\!\!\!\!\!\!\!
&&
\int\!\!\! dx\overline{\xi_n(x{\scriptstyle2n{+}1\over|\lambda|})\!}\!
\Bigg[\!{-}\xi''_p(x{\scriptstyle2p+1\over|\lambda|})
{+}{|\lambda|\over|x|}\xi_p(x{\scriptstyle2p{+}1\over|\lambda|})\!\Bigg]\!
{-}\!\!\!\int\!\! dx\!\Bigg[\!{-}\overline{\xi''_n
(x{\scriptstyle2n{+}1\over|\lambda|})}{+}{|\lambda|\over|x|}
\overline{\xi_n(x{\scriptstyle2n{+}1\over|\lambda|}\!)}\Bigg]
\!\xi_p(x{\scriptstyle2p+1\over|\lambda|}\!)\\
\!\!\!\!\!\!\!\!\!\!\!\!\!\!\!\!\!
\!\!\!\!\!\!\!\!\!\!\!\!\!\!\!\!\!\!\!\!\!\!\!\!\!\!\!
&&\qquad\qquad
=\lim_{\varepsilon\to 0^+}|\lambda|\left[{\overline{\xi_n(u)}\over2p+1}
{d\xi_p\over du}(u)-\overline{{d\xi_n\over du}(u)}
{\xi_p(u)\over2n+1}\right]_{-\varepsilon}^\varepsilon
\end{eqnarray*}
is not zero, for instance $\Delta_{01}=-{8\over3\pi}$,
$\Delta_{02}={28\over5\pi}$, $\Delta_{03}=-{116\over7\pi}$,
$\Delta_{12}=-{4\over5\pi}$, $\Delta_{13}={23\over7\pi}$,
$\Delta_{23}={27\over14\pi}$, etc.

But the situations are quite different. In the case of the unbound spectrum,
eigenstates must be strictly orthogonal; otherwise, a quantum of a given energy
$E$, coming from the frontiers of the universe and interacting with the system
would not only create particles of the same energy, but of other energies, so
$E$ becomes blurred; but this blurring would spoil into the whole universe,
which is impossible. So, we have discarded this possibility (proving therefore
$T=0$) of a break of hermiticity of $H$.

On the other hand, a bound state of energy $E$ may relax into a coherent state,
thanks to interacting overlaps between non orthogonal eigenstates. Thus, it may
be excited into a free state of different energy, with a certain probability,
which we will examine; yet, this mechanism does not contradict any physical
law, and is possible.

Moreover, $H$ is still an observable : its spectrum is \emph{real}, and
canonical quantization theory is still valid, so a break of hermiticity
restrictedly for $E\in\{\tilde E_n, n\in\N\}$ does not yield any 
contradiction of quantum mechanics, although its exceeds its standard
axiomatic formulation.

\subsubsection{Coherent bound states}

Anomalous bound states are not orthogonal, so they are not stable: the
spontaneous transition $\tilde E_n\to\tilde T_p$ is allowed, without any
interaction term in the Hamiltonian, which contradicts the standard properties
of quantum mechanics. Therefore, a state of energy $\tilde E_n$ is not stable.
However, the transfer probability between two states of energies
$\tilde E_n$ and $\tilde E_p$ is very small and decreases as
$|\tilde E_n-\tilde E_p|$ is increased, so, anomalous states are almost stable,
and their actual energy is only slightly blurred.
In order to calculate stable states, one simply needs to diagonalize the
(infinite) matrix
$M=(\langle\xi_m|\xi_n\rangle)_{m,n}$. $M$ is replaced
by truncated matrix $M^{(N)}$, of size $N\times N$ corresponding to
$0\le m,n\le N-1$, and we have diagonalized $M^{(N)}$ instead. By chance, the
coefficients of $M^{(N)}$ are rapidly converging when $N$ is increased, so we
can calculate numerically those of $M$.
 
\begin{figure}%[-h]
\centering
\includegraphics[width=6cm]{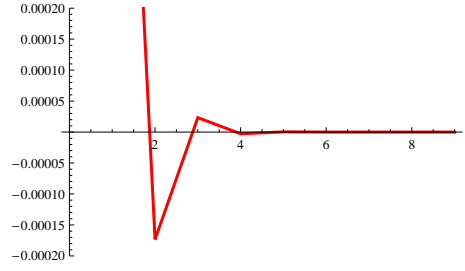}
\caption{$P^{(N)}_{1,N}$ versus $N$.}
\label{deviation}
\end{figure}

\begin{figure}%[-h]
\centering
\includegraphics[width=6cm]{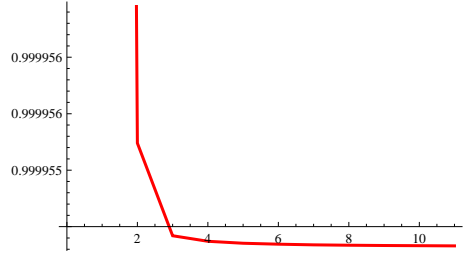}
\caption{$M^{(N)}_{1,1}$ versus $N$ (it is normalized to 1).}
\label{deviation1}
\end{figure}

\begin{figure}%[-h]
\centering
\includegraphics[width=7cm]{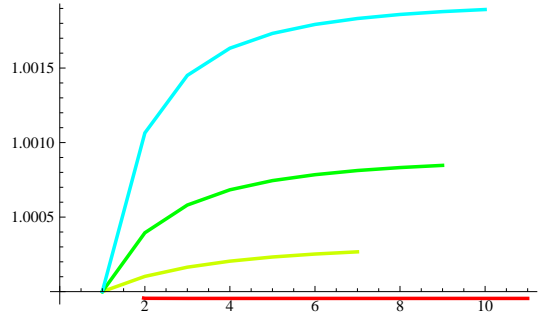}
\caption{$M^{(N)}_{1,1}$ (red), $M^{(N)}_{1,2}$ (blue),
$M^{(N)}_{1,3}$ (green) and $M^{(N)}_{1,5}$ (yellow) versus $N$ (coefficient
$P^{(N)}_{1,q}$ is divided by $P^{(q)}_{1,q}$ to show the relative
convergence).}
\label{convergence}
\end{figure}

Let $P^{(N)}$ be the corresponding change of basis matrix. $P^{(N)}$ is indeed
close to  unity; we show, in Fig.~\ref{deviation} the rapid decrease of
$P^{(N)}_{1,N}$ versus $N$, in Fig.~\ref{deviation1} the diagonal coefficient
$P^{(N)}_{1,1}$ versus $N$, and, in Fig.~\ref{convergence} the convergence
of $P^{(N)}_{1,q}$ versus $N$, for some values of $q$ (these coefficients are
divided by $P^{(q)}_{1,q}$ for convenience). One verifies that the
diagonal coefficient deviation from 1 remains very small, and, correspondingly,
that other coefficients are of several orders smaller.

The stable states that we have calculated are coherent states. Each coherent
state can be labeled by the closest state of energy $\tilde E_n$  and will be
written $\tilde\xi_n$. When a state of energy $\tilde E_n$ is created, it
will relax to $\tilde\xi_n$. The delay of this relaxation is of the
order $\hbar\over\Delta\tilde E_n$, where $\Delta\tilde E_n$ is the uncertainty
of $\tilde E_n$ due to the instability process and can be explicitly
calculated.

On the other hand, consider an excited state of energy $E=-\tilde E_p$~; even if
the state was initially created as $\xi_n$ with $n\ne p$, the
probability of exciting state $\xi_p$, although small, is never zero.

\subsubsection{Orthogonality between regular and anomalous states}

Finally, we would like to insist on the orthogonality between regular and
anomalous states. Otherwise, spontaneous relaxation between regular states,
$E_n\to E_p$, might occur, through channel $E_n\to\tilde E_q\to E_p$, and the
effective overlap between regular states would not be zero.

If one adds, in the Hamiltonian, an interaction term between regular and
anomalous term, allowing in-between transitions, the exact calculation of
transfer probability would become more complicated, because of the relaxation
process.

Eventually, in a real system, one should take into account the dynamical aspect
of the problem, and consider, instead of a coherent state, an intermediate
state, which would include the real dynamical relaxation process. Although it
may seem complicated, this opens interecting directions.

\section{Conclusion}

Simple quantum mechanics can always bring new and surprising results. Indeed, we
have found that the hermiticity of the Coulomb Hamiltonian may break
exclusively for a closed family of bound states, which we therefore called
anomalous states. These states are not stable, and one can only observe,
instead, coherent states. We have also found a new case of quantum reflection,
by solving the one dimension Coulomb problem.
\\[0.2cm]
\noindent
{\bf Acknowledgments} \\
We would like to thank J.-M. Luck and R. Balian for invaluable help and
suggestions.
\appendix
\section*{Appendix}
\setcounter{equation}{35}
\setcounter{figure}{8}
\renewcommand{\theequation}{\arabic{equation}}
\def\thesubsection{\Alph{subsection}}
\subsection{Asymptotic behavior of $F_\eta$ and $G_\eta$ when $u\to-\infty$}
\label{asymptotic}

Here we analyze the asymptotic behavior of $F_\eta(u)$ and $G_\eta(u)$ for
$u \to -\infty$. Our results are different from those in Eqs.~(6c) and (6d) in
Ref.~\cite{Mineev}, (see \cite{mistake}). 

Let us first demonstrate (\ref{compFmoins}). First note that 
\begin{equation}
\label{symF}
t\e^{\iii t}M(1+\ii\eta,2,-2\ii u)=
\overline{t\e^{-\iii t}M(1-\ii\eta,2,2\ii u)}\ ,
\end{equation}
but, since it is real, one can omit the conjugation. For $u>0$,
writing $u=|u|$ and using~(\ref{compFplus}), one gets
$$
|u|\e^{-\iii |u|}M(1-\ii \eta,2,2\ii |u|)\largesim{|u|\to+\infty}
\e^{\pi\eta\over2}\kappa_\eta\sin(|u|-\Theta_\eta(u))\ .
 $$
For $u<0$, writing $u=-|u|$ and using~(\ref{symF}), one gets
$$
|u|\e^{\iii |u|}M(1+\ii \eta,2,-2\ii|u|)\largesim{|u|\to+\infty}
\e^{\pi\eta\over2}
\kappa_\eta\underbrace{\sin(|u|-\Theta_\eta(u))}_{=-\sin(u+\Theta_\eta(u))}\ ;
 $$
if you make $\eta\to-\eta$ in the last relation, and multiply by -1, you get
$$
-|u|\e^{\iii |u|}M(1-\ii \eta,2,-2\ii|u|)\largesim{|u|\to+\infty}
\e^{-{\pi\eta\over2}}\kappa_\eta\sin(u-\Theta_\eta(u))\ ,
 $$
which is exactly the expected relation
$$
F_\eta(-|u|)\largesim{|u|\to+\infty}
\e^{-\pi\eta}\sin(u-\Theta_\eta(u))\ .
 $$
We only write here the leading order of (\ref{compFmoins}), you must be
very careful of all sign compensations for the next orders. Eventually,
if one makes again $\eta\to-\eta$ in the last relation, one gets directly
$$
F_{-\eta}(-|u|)\largesim{|u|\to+\infty}
\e^{\pi\eta}\sin(u+\Theta_\eta(u))\ ,
 $$
which is the behavior of $f_\eta(u)$ for $u\sim-\infty$.

The demonstration is very similar, for (\ref{compGmoins}). First note that
\begin{equation}
\label{symG}
t\e^{\iii t}U(1+\ii\eta,2,-2\ii u)=
\overline{t\e^{-\iii t}U(1-\ii\eta,2,2\ii u)}\ ;
\end{equation}
here, conjugation can not be omitted. For $u>0$, writing $u=|u|$,
using~(\ref{compGplus}) and keeping only the real part, one gets
$$
\Re\left(|u|\e^{-\iii |u|}U(1-\ii \eta,2,2\ii |u|)\right)
\largesim{|u|\to+\infty}
{\e^{-{\pi\eta\over2}}\over2\eta\Re(\Gamma(-\ii\eta))}
{\cos(|u|-\Theta_\eta(u))\over\kappa_\eta}\ .
 $$
For $u<0$, writing $u=-|u|$, using~(\ref{symG}) and still keeping only the
real part, one gets
$$
\Re\left(|u|\e^{\iii |u|}U(1+\ii \eta,2,-2\ii|u|)\right)
\largesim{|u|\to+\infty}
{\e^{-{\pi\eta\over2}}\over2\eta\Re(\Gamma(\ii\eta))}{1\over\kappa_\eta}
\underbrace{\cos(|u|-\Theta_\eta(u))}_{={\cos(u+\Theta_\eta(u))}}\ .
 $$
if you make $\eta\to-\eta$ in the last relation, and multiply by -1, you get
$$
\Re\left(-|u|\e^{\iii |u|}U(1-\ii \eta,2,-2\ii|u|)\right)
\largesim{|u|\to+\infty}{\e^{\pi\eta\over2}\over2\eta\Re(\Gamma(-\ii\eta))}
{\cos(u-\Theta_\eta(u))\over\kappa_\eta}\ ,
 $$
which is exactly
$$
G_\eta(-|u|)\largesim{|u|\to+\infty}\e^{\pi\eta}\cos(u-\Theta_\eta(u))\ .
 $$ 
Eventually, if one makes again $\eta\to-\eta$ in the last relation, one gets
directly
$$
G_{-\eta}(-|u|)\largesim{|u|\to+\infty}
\e^{-\pi\eta}\cos(u+\Theta_\eta(u))\ ,
 $$
which is the behavior of $g\eta(u)$ for $u\sim-\infty$.

\subsection{Expression of $t$ as a function of $T$}
\label{Texpression}

First, you get simple relations between $(t_\alpha,r_\alpha)$ and
$(A_\alpha,B_\alpha,a_\alpha,b_\alpha)$ ($\alpha=$R,L)~:
\begin{subeq}
\begin{eqnarray}
A_{\rm L}&=&\ii t_{\rm L}\ ;\label{AL}\\
B_{\rm L}&=&t_{\rm L}\ ;\label{BL}\\
a_{\rm L}&=&\ii\e^{-\pi\eta}(1-r_{\rm L})\ ;\label{aL}\\
b_{\rm L}&=&\e^{\pi\eta}(1+r_{\rm L})\ ;\label{bL}\\
A_{\rm R}&=&-\ii(1-r_{\rm R})\ ;\label{AR}\\
B_{\rm R}&=&1+r_{\rm R}\ ;\label{BR}\\
a_{\rm R}&=&-\ii\e^{-\pi\eta} t_{\rm R}\ ;\label{aR}\\
b_{\rm R}&=&\e^{\pi\eta}t_{\rm R}\ .\label{bR}
\end{eqnarray}
\end{subeq}

The unitarity of $S$ writes
\begin{subeq}
\begin{eqnarray}
\label{RT1}
|r|^2+|t|^2&=&1\ ;\\
\overline{t}r+r\overline{t}&=&0\ .
\label{rtnul}
\end{eqnarray}
\end{subeq}

From (\ref{rtnul}) one deduce 
\begin{equation}
{t\over|t|}=\ii\epsilon{r\over|r|}\ ,
\label{epsilon}
\end{equation} where $\epsilon=\pm1$. 
From relations~(\ref{BL},\ref{bL}), one gets
$$
{b_{\rm L}\e^{-\pi\eta}\over B_{\rm L}}={1+r\over t}\ .
 $$
By use of relations~(\ref{RT1},\ref{rtnul}), this writes
$$
{b_{\rm L}\e^{-\pi\eta}\over B_{\rm L}}=
{1+\ii\epsilon t\sqrt{1-T\over T}\over t}\ ,
 $$
but Eq. (\ref{continu}) implies the existence of $\theta\in\R$ such that
$$
{b_{\rm L}\e^{-\pi\eta}\over B_{\rm L}}=\e^{\iii\theta}\ ,
 $$
so, using back relation (\ref{defRT}), we get
$$
{1\over t}=\e^{\iii\theta}-\ii\epsilon \sqrt{{1\over|t|^2}-1}\ .
 $$ 
We carefully multiply this equation by its conjugate and find
$$
{1\over|t|^2}=1+{1\over|t|^2}-1-2\epsilon\sin(\theta)\sqrt{{1\over|t|^2}-1}\ ,
 $$
which implies $\theta=0$ or $\pi$. We will write $\e^{\iii\theta}=\epsilon'$
then
$$
{1\over t}-\epsilon'=-\ii\epsilon \sqrt{{1\over|t|^2}-1}\ .
 $$
We carefully multiply this equation by its conjugate and find
$$
{1\over|t|^2}+1-\epsilon'{2\Re(t)\over|t|^2}={1\over|t|^2}-1
\iff \Re(t)=\epsilon'|t|^2\ ,
 $$
but $|t|^2=\Re(t)^2+\Im(t)^2$, so we get
$$
|t|^2=|t|^4+\Im(t)^2\iff \Im(t)=\epsilon''\sqrt{|t|^2-|t|^4}\ .
 $$
By use of (\ref{defRT}), we have
$t=\Re(t)+\ii\Im(t)=\epsilon'T+\ii\epsilon''\sqrt{T-T^2}$. We eventually shall
prove that $\epsilon''=\epsilon$. We put the last expression of $t$ into
$(1+r)/t$ and get
\begin{eqnarray*}
\!\!\!\!\!\!\!\!\!\!\!\!\!\!\!\!\!\!\!\!\!\!\!\!\!\!\!
{1+r\over t}&=&{1+\ii\epsilon t\sqrt{{1\over T}-1}\over t}
={(1-\epsilon\epsilon''+T\epsilon(\epsilon''+\ii\epsilon'\sqrt{{1\over T}-1}))
(\epsilon'T-\ii\epsilon''\sqrt{T-T^2})\over T}\\
\!\!\!\!\!\!\!\!\!\!\!\!\!\!\!\!\!\!\!\!\!\!\!\!\!\!\!
&=&\epsilon'+\ii(\epsilon-\epsilon'')\sqrt{{1\over T}-1}\ .
\end{eqnarray*}
By taking the modulus of this expression, you would find indeed that
$\epsilon=\epsilon''$. However, we already know that it is real
(because $\theta=0$ or $\pi$), so you have the result straight. Now, if you use
back the different relations, you can get the final expression of $T$~:
\begin{equation}
t=\epsilon' T+\ii\epsilon\sqrt{T(1-T)}
\label{tT}
\end{equation}
where  $\epsilon'=\pm1$ is independent of $\epsilon$. By use of relations
(\ref{AR},\ref{BR},\ref{aR},\ref{bR},\ref{AL},\ref{BL},\ref{aL},\ref{bL}), 
(\ref{tT}) and (\ref{continu}), after some calculations, one gets
\begin{subeq}
\begin{eqnarray}
A_{\rm L}&=&-\epsilon\sqrt{T(1-T)}+\ii\epsilon' T\ ;\label{Al}\\
B_{\rm L}&=&\epsilon' T+\ii\epsilon\sqrt{T(1-T)}\ ;\label{Bl}\\
a_{\rm L}&=&\e^{-\pi\eta}(\epsilon\epsilon'\sqrt{T(1-T)}+\ii(2-T))\ ;
\label{al}\\
b_{\rm L}&=&\e^{\pi\eta}(T+\ii\epsilon\epsilon'\sqrt{T(1-T)})\ ;\label{bl}\\
A_{\rm R}&=&-\epsilon\epsilon'\sqrt{T(1-T)}-\ii(2-T)\ ;\label{Ar}\\
B_{\rm R}&=&T+\ii\epsilon\epsilon'\sqrt{T(1-T)}\ ;\label{Br}\\
a_{\rm R}&=&\e^{-\pi\eta}(\epsilon\sqrt{T(1-T)}-\ii\epsilon' T)\ ;\label{ar}\\
b_{\rm R}&=&\e^{\pi\eta}(\epsilon' T+\ii\epsilon\sqrt{T(1-T)})\ ;\label{br}
\end{eqnarray}
and
\begin{equation}
r=T-1+\ii\epsilon\epsilon'\sqrt{T(1-T)}\ .\label{rT}\\
\end{equation}
\end{subeq}
Using these relations, one verifies all
relations~(\ref{defRT},\ref{RT1},\ref{rtnul}) and (\ref{continu2}). 

An important collateral result from this demonstration is indeed that
$$
{b_{\rm L}\e^{-\pi\eta}\over B_{\rm L}}=\epsilon'\ ;
 $$
from relations (\ref{BR},\ref{BL},\ref{bR},\ref{bL}), one gets
$$
{b_{\rm R}\e^{-\pi\eta}\over B_{\rm R}}=
{b_{\rm L}\e^{-\pi\eta}\over B_{\rm L}}=\epsilon'
 $$
which proves, by linearity, relation (\ref{continu2}).

\subsection{Mclaurin expansions}
\label{mclaurin}

Here we study the behavior of basic solutions $f_\eta(u)$, $g_\eta(u)$ and their
derivatives when $u\to0$. Let us consider first the expansion of $F_\eta$ and
$G_\eta$ for $u\to 0^+$, which are given by\cite{Abramowitz}~:
\begin{eqnarray*}
\!\!\!\!\!\!\!\!\!\!\!\!\!\!\!\!\!\!\!\!\!\!\!\!
F_\eta(u)&\simeq&\e^{-{\pi\eta\over2}}|\Gamma(1+\ii\eta)|(u+\eta t^2)\nonumber\\
\!\!\!\!\!\!\!\!\!\!\!\!\!\!\!\!\!\!\!\!\!\!\!\!
&=&C_\eta(u+\eta t^2)\\
\!\!\!\!\!\!\!\!\!\!\!\!\!\!\!\!\!\!\!\!\!\!\!\!
G_\eta(u)&\simeq&{1\over C_\eta}\left\{
2\eta(u+\eta u^2)\Big(\log(2u)-1
+p(\eta)+2\gamma_{\rm E}\big)+(1-{1+6\eta^2\over2}u^2)\right\}
\end{eqnarray*}
\begin{eqnarray*}
\!\!\!\!\!\!\!\!\!\!\!\!\!\!\!\!\!\!\!\!\!\!\!\!
{dF_\eta\over du}(u)&\simeq&C_\eta(1+2\eta u)\\
\!\!\!\!\!\!\!\!\!\!\!\!\!\!\!\!\!\!\!\!\!\!\!\!
{dG_\eta\over du}(u)&\simeq&{1\over C_\eta}\left\{
2\eta\left[(1+2\eta u)\Big(\log(2u)+p(\eta)
+2\gamma_{\rm E}\Big)-\eta u\right]-(1+6\eta^2)u\right\}
\end{eqnarray*}
\begin{eqnarray*}
\!\!\!\!\!\!\!\!\!\!\!\!\!\!\!\!\!\!\!\!\!\!\!\!
{d^2F_\eta\over du^2}(u)&\simeq&C_\eta2\eta\\
\!\!\!\!\!\!\!\!\!\!\!\!\!\!\!\!\!\!\!\!\!\!\!\!
{d^2G_\eta\over du^2}(u)&\simeq&{1\over C_\eta}\left\{
2\eta\left[2\eta\Big(\log(2u)+p(\eta)+2\gamma_{\rm E}\Big)
+\eta+{1\over u}\right]-(1+6\eta^2)\right\}
\end{eqnarray*}
with $p(\eta)=\Re\left(\scriptstyle
{\Gamma'(1+\iii\eta)\over\Gamma(1+\iii\eta)}\right)=p(-\eta)$ and
$\gamma_{\rm E}$ is Euler's constant.
Thus, one gets, at first order, for the complete solution $\varphi$,
\begin{subeq}
\begin{eqnarray}
\!\!\!\!\!\!\!\!\!\!\!\!\!\!\!\!\!\!\!\!\!\!\!\!
\varphi(u,\eta)&\largesim{u\to 0^+}&
B{1\over C_\eta}\ ;
\label{Continue}\\
\!\!\!\!\!\!\!\!\!\!\!\!\!\!\!\!\!\!\!\!\!\!\!\!
\varphi(u,\eta)&\largesim{u\to 0^-}&
b{\e^{-\pi\eta}\over c_\eta}={b\over C_{-\eta}}\ ;
\label{continue}\\
\!\!\!\!\!\!\!\!\!\!\!\!\!\!\!\!\!\!\!\!\!\!\!\!
{\partial\varphi\over\partial u}(u,\eta)&\largesim{u\to 0^+}&
AC_{-\eta}+2B\eta{1\over C_{-\eta}}(\log(2u)+p(\eta)+2\gamma_{\rm E})\ ;
\label{Signe}\\
\!\!\!\!\!\!\!\!\!\!\!\!\!\!\!\!\!\!\!\!\!\!\!\!
{\partial\varphi\over\partial u}(u,\eta)&\largesim{u\to 0^-}&
aC_{-\eta}-2b\eta{1\over C_{-\eta}}(\log(-2u)+p(\eta)+2\gamma_{\rm E})\ .
\label{signe}
\end{eqnarray}
\end{subeq}

\subsection{Orthonormality relations}
\label{orthonormalisation}

The purpose of this section is to calculate the limit, when $L\to\infty$ of
$\int_{-L}^L\overline{\psi(x,E_1,\alpha_1)}\psi(x,E_2,\alpha_2)dx$. Consider a
given $L$, this integral with all functions replaced by their
asymptote~(\ref{pinfty}) or (\ref{ninfty}) becomes:
\begin{eqnarray*}
\!\!\!\!\!\!\!\!\!\!\!\!\!\!\!\!\!\!\!\!\!\!\!\!
{1\over2}\int_0^L\!\!\!\!\!\!\!\!\!\!\!\!\!\!\!\!\!\!&dx&
\cos\left({\lambda x\over2{\eta_1\eta_2\over\eta_1-\eta_2}}
+\Theta_{\eta_1}({x\lambda\over2\eta_1})-
\Theta_{\eta_2}({x\lambda\over2\eta_2})\right)A^+_{\alpha_1\alpha_2}\\
\!\!\!\!\!\!\!\!\!\!\!\!\!\!\!\!\!\!\!\!\!\!\!\!
&-&\cos\left({\lambda x\over2{\eta_1\eta_2\over\eta_1+\eta_2}}
-\Theta_{\eta_1}({x\lambda\over2\eta_1})-
\Theta_{\eta_2}({x\lambda\over2\eta_2})\right)A^-_{\alpha_1\alpha_2}\\
\!\!\!\!\!\!\!\!\!\!\!\!\!\!\!\!\!\!\!\!\!\!\!\!
&+&\sin\left({\lambda x\over2{\eta_1\eta_2\over\eta_1+\eta_2}}
-\Theta_{\eta_1}({x\lambda\over2\eta_1})-
\Theta_{\eta_2}({x\lambda\over2\eta_2})\right)B^+_{\alpha_1\alpha_2}\\
\!\!\!\!\!\!\!\!\!\!\!\!\!\!\!\!\!\!\!\!\!\!\!\!
&+&\sin\left({\lambda x\over2{\eta_1\eta_2\over\eta_1-\eta_2}}
+\Theta_{\eta_1}({x\lambda\over2\eta_1})-
\Theta_{\eta_2}({x\lambda\over2\eta_2})\right)B^-_{\alpha_1\alpha_2}\\
\!\!\!\!\!\!\!\!\!\!\!\!\!\!\!\!\!\!\!\!\!\!\!\!
+{1\over2}\int_{-L}^0\!\!&dx&
\cos\left({\lambda x\over2{\eta_1\eta_2\over\eta_1-\eta_2}}
-\Theta_{\eta_1}({x\lambda\over2\eta_1})+
\Theta_{\eta_2}({x\lambda\over2\eta_2})\right)a^+_{\alpha_1\alpha_2}\\
\!\!\!\!\!\!\!\!\!\!\!\!\!\!\!\!\!\!\!\!\!\!\!\!
&-&\cos\!\!\left(\!\!{\lambda x\over2{\eta_1\eta_2\over\eta_1+\eta_2}}
+\Theta_{\eta_1}({x\lambda\over2\eta_1})+
\Theta_{\eta_2}({x\lambda\over2\eta_2})\right)a^-_{\alpha_1\alpha_2}\\
\!\!\!\!\!\!\!\!\!\!\!\!\!\!\!\!\!\!\!\!\!\!\!\!
\!\!\!\!\!\!\!\!\!\!\!\!\!\!\!\!\!\!\!\!\!\!\!\!
&+&\sin\!\!\left(\!\!{\lambda x\over2{\eta_1\eta_2\over\eta_1+\eta_2}}
+\Theta_{\eta_1}({x\lambda\over2\eta_1})+
\Theta_{\eta_2}({x\lambda\over2\eta_2})\right)b^+_{\alpha_1\alpha_2}\\
\!\!\!\!\!\!\!\!\!\!\!\!\!\!\!\!\!\!\!\!\!\!\!\!
&+&\sin\!\!\left(\!\!{\lambda x\over2{\eta_1\eta_2\over\eta_1-\eta_2}}
-\Theta_{\eta_1}({x\lambda\over2\eta_1})+
\Theta_{\eta_2}({x\lambda\over2\eta_2})\right)b^-_{\alpha_1\alpha_2}
\end{eqnarray*}
where we use
\begin{eqnarray*}
A^+_{\alpha_1\alpha_2}=
(\overline{A_{\alpha_1}}A_{\alpha_2}+\overline{B_{\alpha_1}}B_{\alpha_2})
\ ;\qquad
A^-_{\alpha_1\alpha_2}=
(\overline{A_{\alpha_1}}A_{\alpha_2}-\overline{B_{\alpha_1}}B_{\alpha_2})\ ;\\
B^+_{\alpha_1\alpha_2}=
(\overline{A_{\alpha_1}}B_{\alpha_2}+\overline{B_{\alpha_1}}A_{\alpha_2})
\ ;\qquad
B^-_{\alpha_1\alpha_2}=
(\overline{A_{\alpha_1}}B_{\alpha_2}-\overline{B_{\alpha_1}}A_{\alpha_2})\ ;
\end{eqnarray*}
\vglue-0.5cm
\begin{eqnarray*}
a^+_{\alpha_1\alpha_2}&=&
\overline{a_{\alpha_1}}a_{\alpha_2}\e^{\pi(\eta_1+\eta_2)}
+\overline{b_{\alpha_1}}b_{\alpha_2}\e^{-\pi(\eta_1+\eta_2)})\ ;\\
a^-_{\alpha_1\alpha_2}&=&
\overline{a_{\alpha_1}}a_{\alpha_2}\e^{\pi(\eta_1+\eta_2)}
-\overline{b_{\alpha_1}}b_{\alpha_2}\e^{-\pi(\eta_1+\eta_2)})\ ;\\
b^+_{\alpha_1\alpha_2}&=&
\overline{a_{\alpha_1}}b_{\alpha_2}\e^{\pi(\eta_1-\eta_2)}
+\overline{b_{\alpha_1}}a_{\alpha_2}\e^{-\pi(\eta_1-\eta_2)})\ ;\\
b^-_{\alpha_1\alpha_2}&=&
\overline{a_{\alpha_1}}b_{\alpha_2}\e^{\pi(\eta_1-\eta_2)}
-\overline{b_{\alpha_1}}a_{\alpha_2}\e^{-\pi(\eta_1-\eta_2)})\ .
\end{eqnarray*}

The difference with the exact limit is finite and contributes to
constant $c$ in formula~(\ref{orthoL}). Now, these integrations are easily
performed when one notes that all $\Theta_\eta(u)$ functions can be treated as
constant. Indeed, let us consider a simpler integral
$\int_0^L\cos(s u+\ln(u))du$, where we will omit the problem at $u=0$, and
$\delta(L)\equiv
{1\over s}\sin(s u+\ln(u))-\int_0^L\cos(s u+\ln(u))du$ is the
difference of the approximate integral with the exact one. Then,
$\delta'(L)={\sin(s L+\ln(L))\over s L}$ not only tends to zero when
$L\to\infty$, but has a finite integral $\int_0^L\delta'(u)du$. This proves that
all such approximations are valid and simply contribute to constant $c$.

The $x=0$ boundary only contributes to constant $c$ (you may need to replace
$x=0$ with another boundary, in order to avoid any divergence, but this
replacement simply gives another contribution to constant $c$) so we may skip it
and eventually get
\begin{eqnarray*}
{1\over\lambda}\Bigg[{\eta_1\eta_2\over\eta_1-\eta_2}
\sin\left({\lambda L\over2{\eta_1\eta_2\over\eta_1-\eta_2}}
+\Theta_{\eta_1}({L\lambda\over2\eta_1})-\Theta_{\eta_2}({L\lambda\over2\eta_2})
\right)A^+_{\alpha_1\alpha_2}\\
+{\eta_1\eta_2\over\eta_1+\eta_2}
\sin\left({\lambda L\over2{\eta_1\eta_2\over\eta_1+\eta_2}}
-\Theta_{\eta_1}({L\lambda\over2\eta_1})-\Theta_{\eta_2}({L\lambda\over2\eta_2})
\right)A^-_{\alpha_1\alpha_2}\\
-{\eta_1\eta_2\over\eta_1+\eta_2}
\cos\left({\lambda L\over2{\eta_1\eta_2\over\eta_1+\eta_2}}
-\Theta_{\eta_1}({L\lambda\over2\eta_1})-\Theta_{\eta_2}({L\lambda\over2\eta_2})
\right)B^+_{\alpha_1\alpha_2}\\
-{\eta_1\eta_2\over\eta_1-\eta_2}
\cos\left({\lambda L\over2{\eta_1\eta_2\over\eta_1-\eta_2}}
+\Theta_{\eta_1}({L\lambda\over2\eta_1})-\Theta_{\eta_2}({L\lambda\over2\eta_2})
\right)B^-_{\alpha_1\alpha_2}\\
-{\eta_1\eta_2\over\eta_1-\eta_2}
\sin\left({\lambda L\over2{\eta_1\eta_2\over\eta_1-\eta_2}}
-\Theta_{\eta_1}({L\lambda\over2\eta_1})+\Theta_{\eta_2}({L\lambda\over2\eta_2})
\right)a^+_{\alpha_1\alpha_2}\\
+{\eta_1\eta_2\over\eta_1+\eta_2}
\sin\left({\lambda L\over2{\eta_1\eta_2\over\eta_1+\eta_2}}
+\Theta_{\eta_1}({L\lambda\over2\eta_1})+\Theta_{\eta_2}({L\lambda\over2\eta_2})
\right)a^-_{\alpha_1\alpha_2}\\
+{\eta_1\eta_2\over\eta_1+\eta_2}
\cos\left({\lambda L\over2{\eta_1\eta_2\over\eta_1+\eta_2}}
+\Theta_{\eta_1}({L\lambda\over2\eta_1})+\Theta_{\eta_2}({L\lambda\over2\eta_2})
\right)b^+_{\alpha_1\alpha_2}\\
+{\eta_1\eta_2\over\eta_1-\eta_2}
\cos\left({\lambda L\over2{\eta_1\eta_2\over\eta_1-\eta_2}}
-\Theta_{\eta_1}({L\lambda\over2\eta_1})+\Theta_{\eta_2}({L\lambda\over2\eta_2})
\right)b^-_{\alpha_1\alpha_2}\Bigg]\ .
\end{eqnarray*}
Now, both limits of $\sin(s L)\over s$ and $\cos(s L)\over s$
when $L\to\infty$ are equal to $\pi\delta(s)$ (with differential
$ds$). The $\ln(u)$ correction has no influence (see Appendix~\ref{hermiticite}).
Then we write $\delta({1\over\eta_2}-{1\over\eta_1})=
\delta({2\over\lambda}(k_1-k_2))={\lambda\over2}\delta(k_1-k_2)$, so we
eventually get factor ${\pi\over\lambda}{\lambda\over2}$. We have forgotten the
exact differential $dk\over2\pi$ in one dimension, and we will include a last
factor 2 which accounts for the equality between the limits of $\int_0^L$ and
$\int_{-L}^0$. Altogether, we get formula~(\ref{orthoL}), with 
the following coefficients of matrix $P$~:
\begin{eqnarray*}
P_{\alpha\alpha'}&=&
{\overline{A_\alpha}A_{\alpha'}+\overline{B_\alpha}B_{\alpha'}
+\overline{a_\alpha}a_{\alpha'}\e^{2\pi\eta}+
\overline{b_\alpha}b_{\alpha'}\e^{-2\pi\eta}\over2}
\end{eqnarray*}
and, with relations
(\ref{Ar},\ref{Br},\ref{ar},\ref{br},\ref{Al},\ref{Bl},\ref{al},\ref{bl}), we
eventually get
$$
P=\left(\matrix{1&0\cr0&1\cr}\right)\ ,
 $$
thus (\ref{ortho}) is verified.

\subsection{Hermiticity relations at infinity}
\label{hermiticite}

The calculation of (\ref{j2inf}) is similar to the previous orthonormality
calculations, although simpler. Here $\alpha=R,L$ for the choice of
$\varphi_\alpha$ and we use the notations of Appendix~\ref{orthonormalisation}. One gets
\begin{eqnarray*}
{\eta_1+\eta_2\over2\eta_1\eta_2}
\sin\left({\lambda L\over2{\eta_1\eta_2\over\eta_1-\eta_2}}
+\Theta_{\eta_1}({L\lambda\over2\eta_1})-\Theta_{\eta_2}({L\lambda\over2\eta_2})
\right)A^+_{\alpha\alpha}\\
-{\eta_1-\eta_2\over2\eta_1\eta_2}
\sin\left({\lambda L\over2{\eta_1\eta_2\over\eta_1+\eta_2}}
-\Theta_{\eta_1}({L\lambda\over2\eta_1})-\Theta_{\eta_2}({L\lambda\over2\eta_2})
\right)A^-_{\alpha\alpha}\\
-{\eta_1-\eta_2\over2\eta_1\eta_2}
\cos\left({\lambda L\over2{\eta_1\eta_2\over\eta_1+\eta_2}}
-\Theta_{\eta_1}({L\lambda\over2\eta_1})-\Theta_{\eta_2}({L\lambda\over2\eta_2})
\right)B^+_{\alpha\alpha}\\
+{\eta_1+\eta_2\over2\eta_1\eta_2}
\cos\left({\lambda L\over2{\eta_1\eta_2\over\eta_1-\eta_2}}
+\Theta_{\eta_1}({L\lambda\over2\eta_1})-\Theta_{\eta_2}({L\lambda\over2\eta_2})
\right)B^-_{\alpha\alpha}\\
-{\eta_1+\eta_2\over2\eta_1\eta_2}
\sin\left({\lambda L\over2{\eta_1\eta_2\over\eta_1-\eta_2}}
-\Theta_{\eta_1}({L\lambda\over2\eta_1})+\Theta_{\eta_2}({L\lambda\over2\eta_2})
\right)a^+_{\alpha\alpha}\\
+{\eta_1-\eta_2\over2\eta_1\eta_2}
\sin\left({\lambda L\over2{\eta_1\eta_2\over\eta_1+\eta_2}}
+\Theta_{\eta_1}({L\lambda\over2\eta_1})+\Theta_{\eta_2}({L\lambda\over2\eta_2})
\right)a^-_{\alpha\alpha}\\
+{\eta_1-\eta_2\over2\eta_1\eta_2}
\cos\left({\lambda L\over2{\eta_1\eta_2\over\eta_1+\eta_2}}
+\Theta_{\eta_1}({L\lambda\over2\eta_1})+\Theta_{\eta_2}({L\lambda\over2\eta_2})
\right)b^+_{\alpha\alpha}\\
-{\eta_1-\eta_2\over2\eta_1\eta_2}
\cos\left({\lambda L\over2{\eta_1\eta_2\over\eta_1-\eta_2}}
-\Theta_{\eta_1}({L\lambda\over2\eta_1})+\Theta_{\eta_2}({L\lambda\over2\eta_2})
\right)b^-_{\alpha\alpha}\ .
\end{eqnarray*}

One striking thing is that the coefficients ${1\over\eta_1}\pm{1\over\eta_2}$
are very different from the previous case. In order to match with the $\delta$
limit, one must divide by ${1\over\eta_1}\mp{1\over\eta_2}$, so there is a
global supplementary factor ${1\over\eta_1}^2-{1\over\eta_2}^2$, which, when
multiplied by $\delta({1\over\eta_1}\pm{1\over\eta_2})$, will always give zero.

Another important difference is that we have made no approximation in this case.
It is worth studying the last limit more carefully, than we did before. Using
again a simpler case, we want to prove that $\lim_{L\to\infty}
{1\over s}\sin(s L-s\ln(L)-\kappa s^2+\beta)$
is $\pi\delta(s)$ ($\kappa$ and $\beta$ are just constants here). The
important thing is that $\tilde L\equiv L-\ln(L)\to\infty$ and can be used as a
parameter, so the result is proved, and the limit of (\ref{j2inf}) is strictly
zero.

\subsection{Digression: To WKB or not to WKB?} 

In a nuclear fission process, a light nucleus of mass $m$ and charge $q=Zq_e>0$
(e.g an alpha particle with $Z=2$) is trapped in a metastable state at energy
$E$ due to a potential ``pocket''  $V(r)=V_N(r)+V_C(r)$ of an heavy nucleus of
charge $q'=Z'q_e$ (here $r$ is the distance between the centers of mass of the
two nuclei). The potential is the sum of a strong short-range attractive nuclear
potential $V_N(r)$ and a repulsive long range Coulomb potential
$V_C(r)=\frac {qq'}{r}$. The focus of interest is on the escape
probability $P$ from the metastable state. In a crude approximation, $V(r)$ is
replaced by a deep potential well of range $R$ and depth $-V_0$ and a Coulomb
tail for $r>R$, see figure~\ref{Gamow1}:
\begin{figure}[!h]
\centering
\includegraphics[width=7cm]{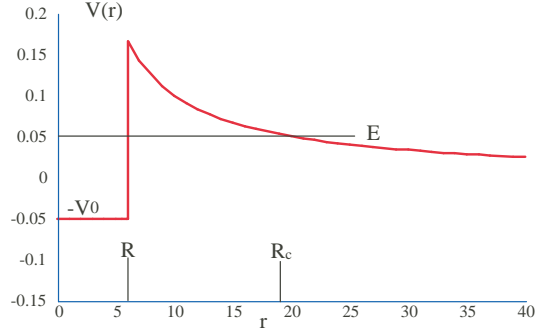}
\caption{ Approximated potential $V(r)=V_N(r)+V_C(r)$ designed to calculate the
fission probability within the WKB approximation.}
\label{Gamow1}
\end{figure}

The escape probability is then calculated in the WKB approximation, integrating
the local momentum $\kappa (r)=\sqrt {\frac{2 m}{\hbar^2}[V_C(r)-E)]}$ between
the turning points $R$ and $R_c$  (such that $V_C(R_c)=E$).
$$
\Lambda=\int_R^{R_c}\!\!\!\! \kappa(r) dr
=\int_0^{R_c}\!\!\!\! \kappa(r) dr
\!-\!\!\int_0^{R}\!\!\kappa(r) dr\equiv \Lambda_G-\Lambda_R\ ;
\qquad P=e^{- 2 \Lambda}\ .
 $$
When $R \ll R_c$ the result is written as,
\begin{equation}
P=e^{-\frac{2 \pi m qq'}{\hbar v}}e^{\frac{32 m qq' R}{\hbar^2}}
\equiv P_G T_R, \label{Gamow}
\end{equation}
where $v$ is the relative velocity and $P_G$ is the \emph{Gamow factor}, which
contains the energy dependence of the escape probability.
Relation~(\ref{solhalf}), with
$\varepsilon=R$, gives the \emph{exact} escape amplitude $=|t|^{2}$ (for the
special case $V_0=0$ but that can be easily modified). It also shows that the
WKB expression (\ref{Gamow}) cannot be used as $R \to 0$ because it yields a
finite escape probability while the exact result (within the naive model of
figure~\ref{Gamow1}) gives zero escape probability. The reason is that the
conditions for the use of the WKB approximation are not met, strictly speaking.

\subsection{Generalization of recurrence equations}
\label{abramowitz}

We study the changes of relations (14.1) in Ref.~\cite{Abramowitz} for the
bound states ($e<0$), in the case $L=0$. Note first that (14.1.1) is also
changed, it writes now as (\ref{lqpos}).

Relation (14.1.6) writes now (we omit the $L=0$ exponent)
$$
A_1=1\ ;\qquad A_2=\eta\ ;\qquad(k+1)(k+2)A_{k+2}=2\eta A_{k+1}+A_k\ .
 $$
Relation (14.1.14) writes now (with our notations)
$$
L_\eta(u)=2\eta K_\eta(u)(\log(2u)-1+
{\Gamma'(1+\eta)\over\Gamma(1+\eta)}+2\gamma_{\rm E})+\theta_\eta(u)
 $$
with (14.1.17) (relation (14.1.15) is useless here)
$$
\theta_\eta(u)=\sum_{k=0}^\infty a_k u^k
 $$
and relations (14.1.18) to (14.1.20) now become
%\begin{eqnarray*}
$$
a_0=1\ ;\quad  a_1=-1\ ;\quad
(k+1)(k+2)a_{k+2}=2\eta a_{k+1}+a_k-2\eta(2k+3)A_{k+2}\ .
 $$
%\end{eqnarray*}
Eventually, note that new relation (14.1.14) also holds for $u<0$ as soon as you
replace $\log(2u)$ by $\log(-2u)$.

\subsection{Identities between confluent hypergeometric functions an modified
Bessel ones}
\label{demonstration}
We found useful identities between confluent hypergeometric functions 
$M({1\over2}\pm n,2,2t)$ or $U({1\over2}\pm n,2,2t)$ and modified Bessel
functions ${\bf I}_n(t)$ or ${\bf K}_n(t)$, for all $n\in\N$. 

These identities appear to generalize some identity established only for $n=0$
or $n=1$~; indeed, from relations (13.6.3) and (13.6.21) of
Ref.~\cite{Abramowitz}, one shows
\begin{subeq}
\begin{eqnarray}
\e^{-t}M({1\over2},2,2t)&=&{\bf I}_0(t)-{\bf I}_1(t)\ ;\\
\e^{-t}U({1\over2},2,2t)&=&{1\over2\sqrt{\pi}}({\bf K}_0(t)+{\bf K}_1(t))\ ;
\end{eqnarray}
\end{subeq}
Thus, it seems possible to generalize this relations and look for solutions of
Eqs.~(\ref{lqpos}) and (\ref{lqneg}) in the form
\begin{subeq}
\begin{equation}
f_n(t)=t\Big(p_n(t){\bf I}_0(t)-q_n(t){\bf I}_1(t)\Big)
\label{solI}
\end{equation}
or
\begin{equation}
g_n(t)=t\Big(p_n(t){\bf K}_0(t)+q_n(t){\bf K}_1(t)\Big)
\label{solK}
\end{equation}
\end{subeq}
(we took advantage of further relations between the polynomials $(p_n,q_n)$
defined in Eq.~(\ref{solI}) and those defined in Eq.~(\ref{solK}) in order to
save notations.)

Although it works well, it proved more efficient to find directly the recurrence
relations which define $p_n$ and $q_n$. Using relation~(13.4.11) of
Ref.~\cite{Abramowitz} for $M({1\over2}-n,2,2t)$, (13.4.10) for
$M({1\over2}+n,2,2t)$, (13.4.26) for $U({1\over2}-n,2,2t)$ or (13.4.23) for
$U({1\over2}+n,2,2t)$, and making the derivative of Eqs.~(\ref{solI}) and
(\ref{solK}) using ${\bf I}'_0={\bf I}_1$,
${\bf I}'_1(t)={\bf I}_0(t)-{\bf I}_1(t)/t$, ${\bf K}'_0=-{\bf K}_1$ and
${\bf K}'_1(t)=-{\bf K}_0(t)-{\bf K}_1(t)/t$, and fixing $p_0=q_0=1$, one finds,
up to some normalisation factors, 
\begin{subeq}
\begin{eqnarray}
\label{recp}
p_{n+1}(x)&=&(2n+3)p_n(x)+2x(p'_n(x)-p_n(x)-q_n(x))\\
\label{recq}
q_{n+1}(x)&=&(2n+1)q_n(x)+2x(q'_n(x)-p_n(x)-q_n(x))
\end{eqnarray}
\end{subeq}
These definitions have one main advantage : these polynomials are real and have
integer coefficients~; let us write $p_n(x)=\sum_{i=0}^na^n_ix^i$ and
$q_n(x)=\sum_{i=0}^nb^n_ix^i$, we get $a_0^n=(2n+1)!!$, $b_0^n=(2n-1)!!$,
$a_n^n=b_n^n=(-4)^n$.

Eventually, let us fix the normalisation problem (note the symmetry between
$M({1\over2}-n,2,2t)$ and $M({3\over2}+n,2,2t)$ or between $U({1\over2}-n,2,2t)$
and $U({3\over2}+n,2,2t)$ and that $(-1)!!=1$): $\forall n\in\N$,
\begin{subeq}
\begin{eqnarray}
\!\!\!\!\!\!\!\!\!\!\!\!\!\!\!\!\!\!\!\!
\!\!\!\!\!\!\!\!\!\!\!\!\!\!\!\!\!\!\!\!
\e^{-t}M({1\over2}-n,2,2t)&=&
{1\over(2n+1)!!}\big(p_n(t)I_0(t)-q_n(t)I_1(t)\big)\ ;\\
\!\!\!\!\!\!\!\!\!\!\!\!\!\!\!\!\!\!\!\!
\!\!\!\!\!\!\!\!\!\!\!\!\!\!\!\!\!\!\!\!
\label{K}
\e^{-t}U({1\over2}-n,2,2t)&=&
{(-1)^n\over2^{n+1}\sqrt{\pi}}\Big(p_n(t)K_0(t)+q_n(t)K_1(t)\Big)\ ;\\
\!\!\!\!\!\!\!\!\!\!\!\!\!\!\!\!\!\!\!\!
\!\!\!\!\!\!\!\!\!\!\!\!\!\!\!\!\!\!\!\!
\e^{-t}M({3\over2}+n,2,2t)&=&
{1\over(2n+1)!!}\big(p_n(-t)I_0(t)+q_n(-t)I_1(t)\big)\ ;\\
\!\!\!\!\!\!\!\!\!\!\!\!\!\!\!\!\!\!\!\!
\!\!\!\!\!\!\!\!\!\!\!\!\!\!\!\!\!\!\!\!
\e^{-t}U({3\over2}+n,2,2t)&=&
{2^n\over(2n+1)!!(2n-1)!!\sqrt{\pi}}\Big({-}p_n(-t)K_0(t)+\;q_n(-t)K_1(t)\Big).
\end{eqnarray}
\end{subeq}


\begin{thebibliography}{fe}
\bibitem{Mineev2} It is not clear why the analytic continuation avoiding zero
did not give all solutions.
\bibitem{attention} Note that relation (13.1.3) in~\cite{Abramowitz} fails here
so one should use instead (13.1.6).
\bibitem{nonelucide} The origin of this we did not elucidate.
\bibitem{mistake} We believe there is a mistake in the analysis in \S\ after
relation (A18) of Ref. \cite{Mineev}.
\bibitem{nonuniversalite} The representation of (\ref{Smat}) (matrix $S$),
(\ref{orthoL}) (orthogonality constraints) and (\ref{twocurrent}) (hermiticity
constraints) in terms of coefficients $T(\eta)$ are peculiar for the Coulomb
problem discussed here, and not valid for any one-dimensional scattering
problem.
\bibitem{implicite} The essential Coulomb properties are implicitly contained in
these relations.
\bibitem{precisiondelta} More precisely, $\delta(k_1+k_2)$ only contribute to
$\psi(x,0)$, which is not an important matter here. It is however interesting to
note that the complete weight of this state, within $T=0$, is found to be zero,
which is exact.
\bibitem{Mineev3}
Mineev claims that it had been already solved many times, but gives no
references, except of Ref.~\cite{Moshinsky}, which gives not many details.
\end{thebibliography}
\end{document}